\documentclass[a4paper,11pt]{fullverllncs}
\usepackage[left=2.5cm,top=2.5cm,right=2.5cm,centering]{geometry}

\usepackage{graphicx}

\usepackage{amsmath,amssymb}
\usepackage{graphicx}
\usepackage{ascmac}
\usepackage{array}
\usepackage{algpseudocode}
\usepackage{amsfonts}
\usepackage{amssymb}
\usepackage{amsmath}
\usepackage{algorithm, algpseudocode}
\usepackage{multirow}
\usepackage{arydshln}
\usepackage{hhline} 
\usepackage[misc,geometry]{ifsym} 

\usepackage[dvipdfmx, colorlinks]{hyperref, xcolor}
\definecolor{winered}{rgb}{0.5,0,0}
\definecolor{darkblue}{rgb}{0,0,0.5}
\definecolor{darkgreen}{rgb}{0,0.3,0}
\hypersetup{
linkcolor=winered,
citecolor=darkblue,
urlcolor=darkgreen
}

\newcommand{\Adv}{\mathsf{Adv}}

\newcommand{\A}{\mathcal{A}}
\newcommand{\B}{\mathcal{B}}
\newcommand{\C}{\mathcal{C}}

\newcommand{\coCDH}{\mathsf{co\mathchar`-CDH}}

\newcommand{\Z}{\mathbb{Z}}
\newcommand{\G}{\mathbb{G}}

\newcommand{\game}{\mathbf{Game}}

\newcommand{\pk}{\mathsf{pk}}
\newcommand{\sk}{\mathsf{sk}}
\newcommand{\pkADM}{\mathsf{pk}_{\mathsf{Fix}}}
\newcommand{\pkM}{\mathsf{pk}_{\mathsf{Agg}}}
\newcommand{\skADM}{\mathsf{sk}_{\mathsf{Fix}}}
\newcommand{\skM}{\mathsf{sk}_{\mathsf{Agg}}}
\newcommand{\rk}{\mathsf{rk}}
\newcommand{\DID}{\mathsf {DID}}

\newcommand{\ord}{\mathrm{ord}}
\newcommand{\RI}{\mathsf{RI}}
\newcommand{\InRI}{\mathsf{InRI}}
\newcommand{\gk}{{q, \G_1, \G_2, \G_T, e, g_1, g_2}}
\newcommand{\AD}{\mathsf{AD}}

\newcommand{\agg}{\mathsf{agg}}
\newcommand{\MOD}{\mathsf{MOD}}
\newcommand{\ADM}{\mathsf{ADM}}

\newcommand{\Fix}{\mathsf{Fix}}

\newcommand{\KeyGen}{\mathsf{KeyGen}}
\newcommand{\Sign}{\mathsf{Sign}}
\newcommand{\Redact}{\mathsf{Redact}}
\newcommand{\Verify}{\mathsf{Verify}}

\newcommand{\RedInf}{\mathsf{RedInf}}
\newcommand{\ThrRed}{\mathsf{ThrRed}}

\newcommand{\RSS}{{\mathsf{RSS}}}
\newcommand{\tnRSS}{(t,n)\mathchar`-\mathsf{RSS}}
\newcommand{\CES}{{\mathsf{CES}}}

\newcommand{\UftnRSS}{\mathsf{Uf\mathchar`-(t,n)\mathchar`-RSS}}
\newcommand{\PritnRSS}{\mathsf{Priv\mathchar`-(t,n)\mathchar`-RSS}}
\newcommand{\TratnRSS}{\mathsf{Tran\mathchar`-(t,n)\mathchar`-RSS}}
\newcommand{\Corrupt}{\mathsf{Corrupt}}
\newcommand{\LR}{\mathsf{LoRredact}}
\newcommand{\SR}{\mathsf{Sign/Redact}}

\newcommand{\priv}{\mathsf{priv}}
\newcommand{\tran}{\mathsf{tran}}
\newcommand{\uf}{\mathsf{uf}}

\spnewtheorem{assumption}{Assumption}{\bfseries}{\itshape}

\begin{document}



\title{A $t$-out-of-$n$ Redactable Signature Scheme\thanks{This is a full version of a paper \cite{TST19} in Cryptology and Network Security - 18th International Conference (CANS 2019).}} 
\author{Masayuki Tezuka\textsuperscript{(\Letter)} \and Xiangyu Su \and Keisuke Tanaka}
\authorrunning{M.Tezuka et al.}
\institute{Tokyo Institute of Technology, Tokyo, Japan\\
\email{tezuka.m.ac@m.titech.ac.jp}}

\maketitle              
\pagestyle{plain}
\noindent
\makebox[\linewidth]{February 20, 2022}

\begin{abstract}
A redactable signature scheme allows removing parts of a signed message without invalidating the signature.
Currently, the need to prove the validity of digital documents issued by governments and enterprises is increasing. 
However, when disclosing documents, governments and enterprises must remove privacy information concerning individuals.
A redactable signature scheme is useful for such a situation.

In this paper, we introduce the new notion of the $t$-out-of-$n$ redactable signature scheme.
This scheme has a signer, $n$ redactors, a combiner, and a verifier.
The signer designates $n$ redactors and a combiner in advance and generates a signature of a message $M$.
Each redactor decides parts that he or she wants to remove from the message and generates a piece of redaction information.   
The combiner collects pieces of redaction information from all redactors, extracts parts of the message that more than $t$ redactors want to remove, and generate a redacted message.
 
We consider the one-time redaction model which allows redacting signatures generated by the signer only once.
We formalize the one-time redaction $t$-out-of-$n$ redactable signature scheme, define security, and give a construction using the pairing based aggregate signature scheme in the random oracle model.

\end{abstract}

\section{Introduction}
\subsection{Redactable Signatures}
Recently, due to the development of IoT devices, the number of electronic data is steadily increasing.
It is indispensable for future information society to make use of these data.
When we use data, it is important to prove that the data has not been modified in any way.
A digital signature enables a verifier to verify the authenticity of $M$ by checking that $\sigma$ is a legitimate signature on $M$.
However, in our real-world scenario, when we use data, the confidential information should be deleted from the original data.
A digital signature cannot verify the validity of a message with parts of the message removed.

A redactable signature scheme ($\RSS$) is a useful cryptographic scheme for such a situation. 
This scheme consists of a signer, a redactor, and a verifier.
A signer signs a message $M$ with a secret key $\sk$ and generates a valid signature $\sigma$.
A redactor who can become anyone removes some parts of a signed message from $M$, generate a redacted message $M'$, and updates the corresponding signature $\sigma'$ without the secret key $\sk$. 
A verifier still verifies the validity of the signature $\sigma'$ on message $M'$ using $\pk$. 

An idea of a redactable signature scheme was introduced by Steinfeld, Bull, and Zheng \cite{SBZ01} as a content extraction signature scheme $(\CES)$.
This scheme allows generating an extracted signature on selected portions of the signed original document while hiding removed parts of portions. Johnson, Molnar, Song, and Wagner \cite{JMSW02} proposed a redactable signature scheme ($\RSS$) which is similar to a content extraction signature scheme.

\subsubsection*{Security.}
Security of a redactable signature scheme was argued in many works.
Brzuska, Busch, Dagdelen, Fischlin, Franz, Katzenbeisser, Manulis, Onete, Peter, Poettering, and Schr{\"{o}}der \cite{BBDFFKMOPPS10} formalized three security notions of a redactable signature for tree-structured messages in the game-based definition.
\begin{itemize}
\item Unforgeability: Without the secret key $\sk$ it is hard to generate a valid signature $\sigma'$ on a message $M'$ except to redact a signed message $(M, \sigma)$.
\item Privacy: Except for a signer and redactors, it is hard to derive any information about removed parts of the original message $M$ from the redacted message $M'$.
\item Transparency: It is hard to distinguish whether $(M, \sigma)$ directly comes from the signer or has been processed by a redactor.
\end{itemize}
Derler, P{\"{o}}hls, Samelin, and Slamanig \cite{DPSS15} gave a general framework of a redactable signature scheme for arbitrary data structures and defined its security. 

\subsubsection*{Additional Functionalities.}
Following additional functionalities for a redactable signature scheme were proposed.
\begin{itemize}
\item Disclosure control \cite{HHHHMSTY08,IIKO11,IKOST08,IKOST09,MLWW17,MHI06,MHI08,MIMSYTI05,SPBPM12}: Miyazaki, Iwamura, Matsumoto, Sasaki, Yoshiura, Tezuka, and Imai \cite{MIMSYTI05} proposed the disclosure control.
The signer or intermediate redactors can control to prohibit further redactions for parts of the message.
\item Identification of a redactor \cite{IKTY05,IKOTY07}: Izu, Kanaya, Takenaka, and Yoshioka~\cite{IKTY05} proposed the redactable signature scheme called ``Partial Information Assuring Technology for Signature'' (PIATS). 
PIATS allows a verifier to identify the redactor of the signed message.
\item Accountability \cite{PS15}: P{\"{o}}hls and Samelin proposed an accountable redactable signature scheme that allows deriving the accountable party of a signed message.

\item Update and Marge \cite{LLP12,PS14}: Lim, Lee, and Park \cite{LLP12} proposed the redactable signature scheme where a signer can update signature by adding new parts of a message. Moreover, P{\"{o}}hls and Samelin \cite{PS14} proposed the updatable redactable signature scheme that can update a signature and marge signatures derived from the same signer.
\end{itemize}

\subsection{Motivation}
Consider the case where a citizen requests the signed secret document disclosure to the government.
To disclose the secret signed document, the government must remove sensitive data from it.
A decision of deletion for confidential information of a document is performed by multiple officers in the government meeting.

One of the simple solutions is that the signer of the secret document gives the signing key $\sk$ to the meeting chair.
The chair takes a vote on removing sensitive information and removes it from the secret document and signed it using $\sk$.
However, if the meeting chair is malicious, it is risky for the secret document signer to give the meeting chair a signing key $\sk$.
Therefore, the secret document signer wants to avoid giving a signing key $\sk$ to others.

If we try to adapt the original $\RSS$ on this situation, we suffer from the following problem.
$\RSS$ allows anyone to redact message parts and even removes the necessary information.
Moreover,  a malicious chair can redact message parts form the signed document regardless of the decision of the officers.

\subsection{Our Contributions}
We introduce the new notion of $t$-out-of-$n$ redactable signature scheme to overcome this problem.
This scheme is composed of a signer, $n$ redactors, a combiner, and a verifier.
The signer designates $n$ redactors and a combiner, generates a key pair $(\pk, \sk)$ and redactor's secret key $\{\rk[i]\}_{i=1}^n$ and sends $\rk[i]$ to the redactor~$i$.
Then signer decides parts of a message that redaction is allowed, signs the message, and sends its signature to $n$ redactor and a combiner.
Each redactor $i$ selects parts of the signed message that he or she wants to remove, generates a piece of redaction information $\RI_i$, and sends it to the combiner.
The combiner collects all redaction information $\{\RI_i\}_{i=1}^n$, extracts signed message parts which at least $t$ redactors want to remove using $\{\RI_i\}_{i=1}^n$, generates the redactable signature.
The verifier can verify the validities of signatures.

Now, we reconsider applying the $t$-out-of-$n$ redactable signature scheme to the above redaction problem.
Let the secret document signer be a signer of the $t$-out-of-$n$ redactable signature scheme, officers be redactors, and the meeting chair be a combiner.
The secret document signer does not have to give the signing key $\sk$ to the chair.
Our $t$-out-of-$n$ redactable signature only allows the chair to redact parts of message which at least $t$ officers wants to remove.

We consider the one-time redaction model which allows redacting signed message only one time for each signature and gives the unforgeability, privacy, and transparency security of the $t$-out-of-$n$ redactable signature scheme in the one-time redaction model.
Also, we give a concrete construction of the $t$-out-of-$n$ redactable signature scheme which satisfies the unforgeability, privacy, and transparency security.
   
Our construction is based on the $(t,n)$-Shamir's secret sharing scheme and the redactable signature scheme proposed by Miyazaki, Hanaoka, and Imai \cite{MHI06} which use the aggregate signature scheme proposed by Boneh, Gentry, Lynn, and Shacham \cite{BGLS03} based on the ${\sf BLS}$ signature scheme \cite{BLS01}.
Our technical point is to adapt $(t,n)$-Shamir's secret share scheme and compute Lagrangian interpolation at the exponent part of the group element to reconstruct information for the redaction. 
Security of our scheme is based on the computational $\coCDH$ assumption in the random oracle model.

\subsection{Road Map}
In Section \ref{Prelimina}, first, we review bilinear maps and Shamir's secret sharing scheme.
In Section \ref{introtnRS}, introduce the notion of $t$-out-of-$n$ redactable signature and define its security notions.
In Section~\ref{OurtnRScont}, we give a construction of  a $t$-out-of-$n$ redactable signature and its security proof.
In Section~\ref{concludepaper}, we conclude this paper.
In Appendix, we provide missing materials in \cite{TST19}.

\section{Preliminaries}\label{Prelimina}
Let $1^{\lambda}$ be the security parameter.  A function $f(k)$ is negligible in $k$ if $f(\lambda) \leq 2^{-\omega(\log \lambda)}$.
PPT stands for probabilistic polynomial time.
For strings $m$ and $r$, $|m|$ is the bit length of $m$ and $m||r$ is the concatenation of $m$ and $r$. 
For a finite set $S$, $\#S$ denotes the number of elements in $S$, $s \xleftarrow{\$} S$ denotes choosing an element $s$ from $S$ uniformly at random. 
$y \leftarrow \mathcal{A}(x)$ denotes that an algorithm $\mathcal{A}$ outputs $y$ for an input $x$.

\subsection{Bilinear Maps}
Let ${\cal G}$ be a bilinear group generator that takes as an input a security parameter $1^{\lambda}$ and outputs the descriptions of multiplicative groups $(\gk)$ where $\G_1$, $\G_2$, and $\G_T$ are groups of prime order $q$, $e$ is an efficient, non-degenerating bilinear map $e:\G_1 \times \G_2 \rightarrow \G_T$, $g_1$, and $g_2$ are generators of the group $\G_1$ and $\G_2$ respectively, and $\phi$ is a computable isomorphism from $\G_2$ to $\G_1$ with $\phi(g_2)=g_1$.
\begin{enumerate}
\item Bilinear: for all $u \in \G_1$, $v \in \G_2$ and $a, b \in \Z$, then $e(u^a, v^b) = e(u, v)^{ab}$.
\item Non-degenerate: $e(g_1, g_2) \neq 1_{\G_T}$.
\end{enumerate}

\begin{definition}[Computational co-Diffie-Hellman Problem]
For a groups $\G_1 = \langle g_1 \rangle$, $\G_2 = \langle g_2 \rangle$ of prime order $q$, define $\Adv^{\coCDH}_{\G_1, \G_2, \A}$ of a PPT adversary $\A$ as
\begin{equation*}
    \Adv^{\coCDH}_{\G_1, \G_2, \A} =
    \Pr \left[\A(g_2, g_2^{\alpha}, h) = h^{\alpha} \middle| \alpha \xleftarrow{\$} \Z_q, h \leftarrow \G_1 \right],
\end{equation*}
where the probability is taken prover the randomness of $\A$ and the random selection of $(\alpha, h)$.
The computational co-Diffie-Hlleman $(\coCDH)$ assumption is that for all adversaries $\A$, the $\Adv^{\coCDH}_{\G_1, \G_2, \A}$ is negligible in $\lambda$.
\end{definition}

\subsection{Shamir's Secret Sharing Scheme}
In order to construct a $t$-out-of-$n$ redactable signature scheme, we use the $(t,n)$-Shamir's secret sharing scheme \cite{Sha79}.
The $(t, n)$-secret sharing scheme is composed of a dealer and $n$ users. 
The dealer decides a secret $s$, computes secret shares $\{s_i\}^n_{i=1}$, and gives the secret share $s_i$ to the user $i$.
If any $t$ of $n$ secret shares or more shares are collected, we can reconstruct the secret $s$ from them.
While, with less than $t$ secret shares, we cannot recover the secret $s$.

\subsubsection*{Shamir's Secret Sharing Scheme}
We refer to the $(t, n)$-shamir's secret sharing scheme.
\begin{enumerate}
\item The dealer chooses the secret $s \in \mathbb{Z}$ and sets $a_0 \leftarrow s$.
\item The dealer chooses $a_1, \cdots, a_{t-1} \in \{0, \cdots, p-1\}$ independently at random and gets the polynomial $f(X)=\sum_{i=0}^{t-1}a_iX^i$.
\item The dealer computes $f(i)$, sets $s_i \leftarrow (i, f(i))$, and sends the secret share $s_i$ to the user $i$.
\end{enumerate}
If we collect $t$ or more secret shares, we can reconstruct the secret $s$ by the Lagrange interpolation.
Let $J\subset \{1,\cdots,n\}$ and $|J|=t$. 
If we have secret shares $\{s_j\}_{j\in J}=\{(j,f(j))\}_{j\in J}$, we can compute $s= \sum_{i \in J} \left( f(i) \prod_{j\in J, j\neq i}j(j-i)^{-1}\right)$.

\section{A $t$-out-of-$n$ Redactable Signature Scheme}\label{introtnRS}
We explain the outline of our proposed $t$-out-of-$n$ redactable signature scheme in the one-time redaction model.
A $t$-out-of-$n$ redactable signature scheme in the one-time redaction model $\tnRSS$ is a signature scheme that has a signer, $n$ redactors, a combiner, and a verifier.
The signer designates $n$ redactors and the combiner. 

The signer selects a threshold $t$ and the number of redactors $n$.
Then, he or she runs key generation algorithm and gets $(\pk, \sk, \{\rk[i]\}_{i=1}^n)$. The $\pk$ is published and the redactor's key $\rk[i]$ is sent to the redactor $i$.

The signer signs a message $M$ with an admissible description $\ADM$ which represents parts of the message that redactors cannot remove from the message $M$.
In the processing of the signing, a random document ID ($\DID$) is added to the message $M$, then the signature $\sigma$ is generated.
$(M, \ADM, \DID, \sigma)$ generated by the signer is sent to $n$ redactors and the combiner. 

Each redactor $i$ checks whether $\DID$ has never been seen before. If he or she has seen it, then aborts.
Also, if the signature is invalid, then aborts.
Otherwise, he or she selects parts of the message that he or she wants to remove and makes the redaction information $\RI_i$ and sends it to the combiner.
The protocol works only once for $\DID$ which redactors have not seen before.

The combiner collects pieces of redaction information $\{\RI_i\}_{i=1}^n$.
From $\{\RI_i\}_{i=1}^n$, the combiner extracts parts which at least $t$ redactor want to remove.
Finally, the combiner outputs the redacted message $M'$, $\ADM$, $\DID$, and its updated valid signature $\sigma'$.

The signature is verified by using the signer's public key $\pk$. 
In the verification, it is possible to prove the validity of the $(M, \ADM, \DID, \sigma)$ made by a legitimate signer or redacted by the redaction protocol for that signature while keeping redactors anonymity.

\subsection{A $t$-out-of-$n$ Redactable Signature Scheme for Set}
In this paper, we focus on the $t$-out-of-$n$ redactable signature scheme in the one-time redaction model for set.
In the following, we assume that a message $M$ is a set and use following notations. 
An admissible description $\ADM$ is a set containing all elements which must not be redacted.\\
A modification instruction $\MOD$ is a set containing all elements which a redactor want to redact from $M$.
$\ADM \preceq M$ means that $\ADM$ is a valid description. (i.e.,$\ADM \cap M = \ADM$.)
$\MOD \overset \ADM {\preceq}M$ means that $\MOD$ is valid redaction description respect to $\ADM$ and $M$.
(i.e., $\MOD \cap \ADM = \emptyset \land \MOD \subset M$.) A redaction $M' \overset \MOD {\leftarrow} M$ would be $M' \leftarrow M \backslash \MOD.$
In the following definition, we explicit $\ADM$ and $\DID$ in the syntax. 

\begin{definition}
A $t$-out-of-$n$ redactable signature scheme in the one-time redaction model $\tnRSS$ $\Pi$ is composed of four components $(\KeyGen, \Sign, \Redact, \Verify)$. 
 \paragraph*{$\KeyGen :$} A key generation algorithm is a randomized algorithm that a signer runs. 
 Given a security parameter $1^{\lambda}$, a threshold $t$ and the number of redactors $n$, return a signer's public key $\pk$, a signer's secret key $\sk$, and redactor's secret keys $\{\rk[i]\}_{i=1}^n$. 
  \paragraph*{ $\Sign :$} A signing algorithm is a randomized algorithm that a signer runs. 
 Given a signer's secret key $\sk$, a message $M$ and an admissible description $\ADM$, return a message $M$, an admissible description $\ADM$, a document ID $(\DID)$, and a signature $\sigma$.
  \paragraph*{$\Redact :$} A redact protocol is a 1-round interactive protocol between the combiner and $n$ redactors.
 Each redactor $i$ generates redaction information $\RI_i$ and sends to the combiner.
  The combiner collects all redaction informations $\{\RI_i\}_{i=1}^n$ and finally outputs the redacted signature $(M', \ADM, \DID, \sigma')$. 
  We describe the protocol as follows:
\begin{itemize}
\item Given an input $(M, \ADM, \DID, \sigma)$ from the signer, each redactor $i$ selects a modification instruction ${\sf MOD}_i$ and runs a redact information algorithm $\RedInf$ with $(\pk, \rk[i], M, \ADM, \DID, \allowbreak \sigma, \MOD_i, \mathbb{L}_i^{u-1})$. 
$\mathbb{L}_{i}^{u-1}$ is the list which stores on $\DID$ sent from the signer.
It is used for $t$-th input of the $\RedInf$ by redactor $i$ and $L^{0}=\emptyset$.
In the processing in $\RedInf$, if $\DID$ is previous input to $\RedInf$ then redactor $i$ stop interacting with a combiner.
Otherwise, output the redact information $\RI_i$ and the updated list $\mathbb{L}^{u}_i$.
Each redactor $i$ sends $\RI_i$ to the combiner. 
\item The combiner runs a deterministic threshold redact algorithm $\ThrRed$ with $(\pk, M, \ADM, \DID, \allowbreak \sigma, \{\RI_i\}_{i=1}^n)$ as an input. 
In the algorithm $\ThrRed$, $\MOD$ is derived from $\{\RI_i\}_{i=1}^n$ and it redacts a message $M$ based on $\MOD$.
$\ThrRed$ outputs a redacted message $M'$, $\ADM$, $\DID$ and the updated signature $\sigma'$. 
Finally, the combiner outputs $(M', \ADM, \DID, \sigma')$ as an output of $\Redact$ protocol.
\end{itemize}
 \paragraph*{$\Verify :$} A verification algorithm is a deterministic algorithm.
 Given an input $(\pk, M, \ADM, \DID, \sigma)$, return either 1 (Accept) or 0 (Reject). 
\end{definition}

\subsubsection{Correctness} We require the correctness that all honestly computed and redacted signatures are accepted.

\begin{definition}[Correctness]
A $t$-out-of-$n$ redactable signature scheme in the one-time redaction model $\tnRSS$ $\Pi$ is correct, $\forall \lambda\in \mathbb{N}$, $\forall k \in \mathbb{N}$,\\ 
$\forall M_0^u$, $\forall \ADM^u\preceq M_0^u$, $\forall \MOD^u_i \overset {\ADM^u} {\preceq}M_0^u$ for $u \in [k]$ and $i \in [n]$, \\
$\forall (\pk, \sk, \{\rk [i]\}_{i=1}^n) \leftarrow \KeyGen(1^{\lambda},t, n)$, \\
For $u = 1$ to $k$,
\begin{equation*}
\forall (M_0, \ADM^u, \DID^u, \sigma_0^u) \leftarrow \Sign(\sk, M_0^u, \ADM^u),
\end{equation*}
For $i \in [n]$,
\begin{equation*}
\begin{split}
&(\RI_i^u, \mathbb{L}_i^{u}) \leftarrow \RedInf(\pk, \rk[i], M_0^u, \ADM^u, \DID^u, \sigma_0^u, \MOD_i^u, \mathbb{L}_{i}^{u-1}),\\
&(M_1^u, \ADM^u, \DID, \sigma_1^u) \leftarrow \ThrRed(\pk, M_0^u, \ADM^u, \DID^u, \sigma_0^u, \{\RI_i^u\}_{i=1}^n), 
\end{split}
\end{equation*}
we require the following for all $u \in [k]$:
\begin{itemize}
\item If $\DID^k \notin \mathbb{L}_i^{u-1}$, $\Verify(\pk, M_t^u, \ADM^u, \DID^u, \sigma_b^u) = 1$ for all $b \in \{0, 1\}$.
\item If $\DID^k \in \mathbb{L}_i^{u-1}$, $\Verify(\pk, M_0^u, \ADM^u, \DID^u, \sigma_0^u) = 1$.
\end{itemize}
\end{definition}

\subsection{Security of $t$-out-of-$n$ Redactable Signature Schemes}
We give the security notion of unforgeability, privacy, and transparency for a redactable signature scheme in the one-time redaction model.

\subsubsection{Unforgeability} Unforgeability requires that without a signer's secret key $\sk$, it should be infeasible to compute a valid signature $\sigma'$ on $(M', \ADM, \DID)$ except to redact a signed message $(M, \ADM, \DID, \sigma)$ even if $t-1$ redactors keys are corrupted.

\begin{definition}[Unforgeability]\label{ufdef}
The unforgeability against redactors security of a $t$-out-of-$n$ redactable signature scheme in the one-time redaction model $\tnRSS$ $\Pi$ is defined by the following unforgeability game between a challenger $\C$ and a PPT adversary $\A$.
\begin{enumerate}
\item $\C$ generates key pairs $(\pk, \sk, \{\rk[i]\}_{i=1}^{n})$ running $\KeyGen(1^{\lambda},t,n)$, and gives $\pk$ to an adversary $\A$. 
\item $\A$ is given access (throughout the entire game) to a sign oracle $\mathcal{O}^{\Sign}(\cdot, \cdot)$ such that $\mathcal{O}^{\Sign}(M, \ADM)$, returns $(M, \ADM, \DID, \sigma) \leftarrow \Sign(\sk, M, \ADM)$.
\item $\A$ is given access (throughout the entire game) to a redact oracle $\mathcal{O}^{\Redact}(\cdot, \cdot, \cdot, \cdot, \cdot)$. 
$\mathcal{O}^{\Redact}$ is defined as follows:\\
For an $u$-th query $(M, \ADM, \DID, \sigma, \MOD)$: 
\begin{enumerate}
\item $(\RI_i, \mathbb{L}^{u}_i) \leftarrow \RedInf(\pk, \rk[i], M, \ADM, \DID, \sigma, \MOD, \mathbb{L}^{u-1}_i)$ for $i=1,...,n$.
\item $(M', \ADM, \DID, \sigma') \leftarrow \ThrRed(\pk, M, \ADM, \DID, \sigma, \{ \RI_i\}_{i=1}^n).$
\item Return $(M', \ADM, \DID, \sigma')$.
\end{enumerate} 
\item $\A$ is given up to $t-1$ times access (throughout the entire game) to a corrupt oracle $\mathcal{O}^{\Corrupt}(\cdot)$, where $\mathcal{O}^{\Corrupt}(i)$ outputs a $\rk[i]$ of a redactor $i$.
\item $\A$ outputs $(M^*, \ADM^*, \DID^*, \sigma^*)$.
\end{enumerate}
A $t$-out-of-$n$ redactable signature scheme in the one-time redaction model $\tnRSS$ $\Pi$ satisfies the unforgeability security if for all PPT adversaries $\A$, the advantage $\Adv^{\UftnRSS}_{\Pi, \A} = \Pr[\Verify(\pk, M^*, \ADM^* \DID^*, \sigma^*) = 1 \land (M^*, \ADM^*, \DID^*) \notin (Q_{\Sign}\cup Q_{\Redact})]$ is negligible in $\lambda$.
Here, $q_s$ is the total number of queries to $\mathcal{O}^{\Sign}$, $(M_i, \ADM_i)$ is an $i$-th input for $\mathcal{O}^{\Sign}$, $(M^i, \ADM^i, \DID^i, \sigma^i)$ is an $i$-th output of $\mathcal{O}^{\Sign}$ and $Q_{\Sign}:= \bigcup_{i=1}^{q_s} \{(M^i, \ADM^i, \DID^i)\}$. 
Also, $q_r$ is the total number of queries to $\mathcal{O}^{\Redact}$, $(M^i, \ADM^i, \DID^i, \sigma^i, \MOD^i)$ is an $i$-th input for $\mathcal{O}^{\Redact}$, $(M{'}^{i}, \ADM^{i}, \DID^i, \sigma{'}^{i})$ is an $i$-th output of $\mathcal{O}^{\Redact}$ and $Q_{\Redact}:= \bigcup_{i=1}^{q_r} \{(M{'}^{i}, \ADM^{i}, \DID^i)\}$.
\end{definition}

\subsubsection{Privacy} Privacy requires that except for a signer, $n$ redactors, and a combiner, it is infeasible to derive information on redacted message parts when given a message-$\ADM$-$\DID$-signature pair.

\begin{definition}[Privacy]
The privacy of a $t$-out-of-$n$ redactable signature scheme in the one-time redaction model $\tnRSS$ $\Pi$ is defined by the following weak privacy game between a challenger $\C$ and a PPT adversary $\A$.
\begin{enumerate}
\item $\C$ generates key pairs $(\pk, \sk, \{\rk[i]\}_{i=1}^{n})$ by running ${\KeyGen}(1^{\lambda}, t, n)$, and gives $\pk$ to an adversary $\A$. 
\item $\A$ is given access (throughout the entire game) to a sign oracle $\mathcal{O}^{\Sign}(\cdot, \cdot)$ such that $\mathcal{O}^{\Sign}(M, \ADM)$, returns $(M, \ADM, \DID, \sigma) \leftarrow \Sign(\sk, M, \ADM)$.
\item $\A$ is given access (throughout the entire game) to a redact oracle $\mathcal{O}^{\Redact}(\cdot, \cdot, \cdot, \cdot, \cdot)$. 
$\mathcal{O}^{\Redact}$ is defined as follows:\\
For an $u$-th query $(M, \ADM, \DID, \sigma, \MOD)$:\\ 
Let $w$ be the number of queries to $\mathcal{O}^{\LR}$ when $\A$ makes an $u$-th query to~$\mathcal{O}^{\Redact}$.
\begin{enumerate}
\item $(\RI_i, \mathbb{L}^{u+2w}_i) \leftarrow \RedInf(\pk, \rk[i], M, \ADM, \DID, \sigma, \MOD, \mathbb{L}^{u+2w-1}_i)$ for \\$i=1,...,n$.
\item $(M', \ADM, \DID, \sigma') \leftarrow \ThrRed(\pk, M, \ADM, \DID, \sigma, \{ \RI_i\}_{i=1}^n).$
\item Return $(M', \ADM, \DID, \sigma')$.
\end{enumerate} 

\item $\A$ is given access (throughout the entire game) to a left-or-right redact oracle $\mathcal{O}^{\LR}(\cdot, \cdot, \cdot, \cdot, \cdot, \cdot)$. $\mathcal{O}^{\LR}$ is defined as follows:\\
For an $w$-th query $(M^0, \ADM^0, \MOD^{0}, M^1, \ADM^1, \MOD^{1})$:\\
Let $u$ be the number of queries to $\mathcal{O}^{\Redact}$ when $\A$ makes an $w$-th query to~$\mathcal{O}^{\LR}$.
\begin{enumerate}
\item Compute $(M^c, \ADM^c, \DID^c, \sigma^c) \leftarrow \Sign(\sk, M^c, \ADM^c)$ for $c \in \{0, 1\}$.
\item For $i=1, \cdots n$, compute
\begin{equation*}
\begin{split}
(\RI^0_{i}, \mathbb{L}^{u+2w-1}) &\leftarrow \RedInf(\pk, \rk[i], M^0, \ADM^0, \DID^0, \sigma^0, \MOD^0, \mathbb{L}_i^{u+2w-2}) \\
(\RI^1_{i}, \mathbb{L}^{u+2w}) &\leftarrow \RedInf(\pk, \rk[i], M^1, \ADM^1, \DID^1, \sigma^1, \MOD^1, \mathbb{L}_i^{u+2w-1}).
\end{split}
\end{equation*}
 \item For $i=1,...,n$, compute
\begin{equation*} 
  (M^c{'}, \ADM^c, \DID^c, \sigma^c{'}) \leftarrow \ThrRed(\pk, M^c, \ADM^c,  \DID^c, \sigma^c, \{\RI^c_{i}\}_{i=1}^n).
  \end{equation*}

 \item If $M^0{'} \neq M^1{'} \lor \ADM_0 \neq \ADM_1$, return $\perp$.
 \item Return $(M^b{'}, \ADM^b, \DID^b, \sigma^b{'})$.
  $($ $b$ is chosen by $\C$ in step 1.$)$
\end{enumerate}
\item $\A$ outputs $b^*$.
\end{enumerate}
A $t$-out-of-$n$ redactable signature scheme in the one-time redaction model $\tnRSS$ $\Pi$ satisfies the privacy security if for all PPT adversaries $\A$, the following advantage $\Adv^{\PritnRSS}_{\Pi, \A} = \left| \Pr[b = b^*] - 1/2 \right|$ is negligible in $\lambda$.
\end{definition}

\subsubsection{Transparency} Transparency requires that except for a signer, $n$ redactors, and a combiner, it is infeasible to distinguish whether a  signature directly comes from the signer or has been redacted by redactors.

\begin{definition}[Transparency]
The privacy of a $t$-out-of-$n$ redactable signature scheme in the one-time redaction model $\tnRSS$ $\Pi$ is defined by the following weak privacy game between a challenger $\C$ and a PPT adversary $\mathcal{A}$.
\begin{enumerate}
\item $\C$ chooses a bit $b \xleftarrow{\$} \{0, 1\}$, generates key pairs $(\pk, \sk, \{\rk[i]\}_{i=1}^{n})$ by running ${\KeyGen}(1^{\lambda}, t, n)$, and gives $\pk$ to an adversary $\A$. 
\item $\A$ is given access (throughout the entire game) to a sign oracle $\mathcal{O}^{\Sign}(\cdot, \cdot)$ such that $\mathcal{O}^{\Sign}(M, \ADM)$, returns $(M, \ADM, \DID, \sigma) \leftarrow \Sign(\sk, M, \ADM)$.
\item $\A$ is given access (throughout the entire game) to a redact oracle $\mathcal{O}^{\Redact}(\cdot, \cdot, \cdot, \cdot, \cdot)$. 
$\mathcal{O}^{\Redact}$ is defined as follows:\\
For an $u$-th query $(M, \ADM, \DID, \sigma, \MOD)$:\\ 
Let $w$ be the number of queries to $\mathcal{O}^{\SR}$ when $\A$ makes an $u$-th query to $\mathcal{O}^{\Redact}$.
\begin{enumerate}
\item $(\RI_i, \mathbb{L}^{u+2w}_i) \leftarrow \RedInf(\pk, \rk[i], M, \ADM, \DID, \sigma, \MOD, \mathbb{L}^{u+2w-1}_i)$ for \\$i=1,...,n$.
\item $(M', \ADM, \DID, \sigma') \leftarrow \ThrRed(\pk, M, \ADM, \DID, \sigma, \{ \RI_i\}_{i=1}^n).$
\item Return $(M', \ADM, \DID, \sigma')$.
\end{enumerate} 

\item $\A$ is given access (throughout the entire game) to a sign or redact oracle $\mathcal{O}^{\SR}(\cdot, \cdot, \cdot)$. $\mathcal{O}^{\SR}$ is defined as follows:\\
For an $w$-th query $(M, \ADM, \MOD)$:\\
Let $u$ be the number of queries to $\mathcal{O}^{\Redact}$ when $\A$ makes an $w$-th query to~$\mathcal{O}^{\SR}$.
\begin{enumerate}
 \item Compute $(M, \ADM, \DID_0, \sigma) \leftarrow \Sign(\sk, M, \ADM)$.
 \item For $i=1,\dots n$, compute 
 \begin{equation*}
 (\RI_i, \mathbb{L}_i^{u+2w-1}) \leftarrow \RedInf (\pk, \rk[i], M, \ADM, \DID^0, \sigma, \MOD, \mathbb{L}_i^{u+2w-2}).
 \end{equation*}
 \item Compute $(M', \ADM, \DID^0, \sigma^0) \leftarrow \ThrRed(\pk, M, \ADM, \DID^0, \sigma, \{\RI_{i}\}_{i=1}^n)$.
 \item Compute $(M', \ADM, \DID^1, \sigma^1) \leftarrow \Sign (\sk, M', \ADM)$.
 \item For $i=1,\dots n$, $\mathbb{L}_i^{u+2w} \leftarrow \mathbb{L}_i^{u+2w-1} \cup \{\DID^1\}$.
 \item Return $(M', \ADM, \DID^b, \sigma^b)$.\\
\end{enumerate}
\item $\mathcal{A}$ outputs $b^*$.
\end{enumerate}
A $t$-out-of-$n$ redactable signature scheme in the one-time redaction model $\tnRSS$ $\Pi$ satisfies the transparency security if for all PPT adversaries $\A$, the following advantage $\Adv^{\TratnRSS}_{\Pi \A} = \left| \Pr[b = b^*] - 1/2 \right|$ is negligible in $\lambda$.
\end{definition}

\begin{theorem}\label{u-tra-pri}
If $t$-out-of-$n$ redactable signature scheme in the one-time redaction model $\tnRSS$ $\Pi$ satisfies transparency, then it satisfies privacy.
\end{theorem}
We prove {\bf {Theorem} \ref{u-tra-pri}} in a similar way of \cite{BBDFFKMOPPS10,DPSS15}. 
We will describe the proof of {\bf {Theorem} \ref{u-tra-pri}} in the full version of this paper.

\section{Our $t$-out-of-$n$ Redactable Signature Scheme}\label{OurtnRScont}
In this section, we give a concrete construction of $t$-out-of-$n$ redactable signature scheme in one-time redaction model $\tnRSS$ $\Pi_1$.
Let $\ell, d$ be polynomials in $\lambda$, $(\gk) \leftarrow \mathcal{G}(1^{\lambda})$, $H:\{0,1\}^* \rightarrow \mathbb{G}_1$ a hash function, and $M$ a message having a set data structure (i.e., $M = \{m_1,...,m_{\ell}\})$ and $\# M \leq \ell$.

\paragraph*{$\KeyGen (1^{\lambda}, t, n) :$} Given a security parameter $1^{\lambda}$, a threshold value $t$, and the number of redactors $n$, the PPT algorithm $\KeyGen$ works as follows:
\begin{enumerate}
\item \label{keygenfix} Choose $\tilde{x} \xleftarrow{\$} \Z_q$, compute $\tilde{y} \leftarrow g_2^{\tilde{x}}$, and set $(\pkADM, \skADM) \leftarrow (\tilde{y}, \tilde{x})$.
\item \label{keygenm} Choose $a_0, a_1, \cdots, a_{t-1} \xleftarrow{\$} \Z_q$ independently at random and gets the polynomial $f(X)=\sum_{i=0}^{t-1}a_iX^i$.
\item For $i=0$ to $n$, compute $x_i \leftarrow f(i),\; y_i \leftarrow g_2^{f(i)}$.
\item Set $(\pkM, \skM) \leftarrow (y_0, x_0)$, $\rk[i] \leftarrow (i, x_i)$ for all $i \in [n]$.
\item Set $(\pk, \sk) \leftarrow ((\pkADM, \pkM, t, n), (\skADM, \skM))$.
\item \label{keygenout} Return $(\pk, \sk, \{\rk[i]\}_{i=1}^n)$.
\end{enumerate}

\paragraph*{$\Sign (\sk, M, \ADM) :$} Given a signer's secret key $\sk$, a message $M$, and $\ADM$ (In this scheme, $\ADM$ represent a set containing all blocks which must not be redacted.), the PPT algorithm $\Sign$ works as follows:
\begin{enumerate}
\item Parse $\sk$ as $(\skADM, \skM)$.
\item If $\ADM \npreceq M$, (i.e., $\ADM \cap M \neq \ADM$.) then abort.
\item Choose document ID $\DID \xleftarrow{\$} \{0, 1\}^{d}$.
\item \label{signhashqadm} Compute $h_{\ADM} \leftarrow H(\DID||\ord({\ADM}))$.\\
$\ord(\ADM)$ denotes a lexicographic ordering to the elements in $\ADM$.
\item \label{signhashqm} For $m_j \in M$, compute $h_{m_j} \leftarrow H(\DID||m_j)$.
\item \label{signfix} Compute $\sigma_{\Fix} \leftarrow h_{\ADM}^{\skADM}$.
\item \label{signm}Compute $\sigma_{\ADM} \leftarrow h_{\ADM}^{\skM}$, $\sigma_{m_j} \leftarrow h_{m_j}^{\skM}$ for $m_j \in M$.
\item \label{signagg}Compute $\Sigma_{\agg} \leftarrow \sigma_{\ADM} \cdot \prod_{m_j \in M} \sigma_{m_j}$.
\item Set $\sigma \leftarrow (\sigma_{\Fix}, \Sigma_{\agg})$.
\item Return $(M, \ADM, \DID, \sigma)$.
\end{enumerate}

\paragraph*{$\Redact :$} $\Redact$ is an interactive protocol between the combiner and $n$ redactor. The combiner interacts with the $n$ redactors and finally outputs the redacted signature. 
Given a tuple $(M, \ADM, \DID, \sigma)$ to $n$ redactors and the combiner from the signer, the interactive protocol works as follows:
\begin{enumerate}
\item Each redactor $i$ selects a modifiction instruction $\MOD_i$. 
Let $\mathbb{L}_i$ be the list which stores $\DID$s, $\mathbb{L}_i^0 = \emptyset$, and $\mathbb{L}_i^{u-1}$ the list which used in the input of $u$-th running of the PPT algorithm $\RedInf$ by the redactor $i$.\\
The redactor $i$ runs $\RedInf (\pk, \rk[i], M, \ADM, \DID, \sigma, \MOD_i, \mathbb{L}_i^{u-1})$.

\paragraph*{${\RedInf}(\pk, \rk[i], M, \ADM, \DID, \sigma, \MOD_i, \mathbb{L}_i^{u-1}):$}
\begin{enumerate}
    \item Parse $\pk$ as $(\pkADM, \pkM, t, n)$ and $\sigma$ as $(\sigma_{\Fix}, \Sigma_{\agg})$.
    \item If $\DID \in \mathbb{L}_i^{u-1}$ then abort.
    \item Update $\mathbb{L}_i^u \leftarrow \mathbb{L}_i^{u-1} \cup \{\DID\}$.
    \item Check $\MOD_i \overset{\ADM}{\preceq}M$. (i.e., $\MOD_i \cap \ADM = \emptyset \land \MOD_i \subset M$.)
    \item \label{redinfhashqadm} Compute $h_{\ADM} \leftarrow H(\DID||\ord({\ADM}))$.\\
$\ord(\ADM)$ denotes a lexicographic ordering to the elements in $\ADM$.
    \item \label{redinfhashqm} For $m_j \in M$, compute $h_{m_j} \leftarrow H(\DID||m_j)$.\item If $e(\sigma_{\Fix}, g_2) \neq e(h_{\ADM}, \pkADM)$ then abort.
    \item If $e(\Sigma_{\agg}, g_2) \neq e \left(h_{\ADM}, \pkM \right) \cdot \prod_{m_j \in M}e(h_{m_j}, \pkM)$ then abort.
    \item For $m_j \in \MOD_i$, compute $\RI_{i,m_j} \leftarrow h_{m_j}^{\rk[i]}$.
    \item For $m_j \notin \MOD_i$, set $\RI_{i,m_j} \leftarrow \emptyset.$     
    \item Set a redaction information $\RI_i$ of redactor $i$ as $\RI_i \leftarrow \{\RI_{i,m_j}\}_{m_j \in M}$
    \item Output $(\RI_i, \mathbb{L}_i^{u})$.
\end{enumerate}
      
          For one $\DID$, redactor $i$ runs $\RedInf$ only once. This can be done by introducing a table $\mathbb{L}_{i}$.\\

\item Each redactor $i$ sends $(i, \RI_i)$ to the combiner. 
\item The combiner collects all $n$ redaction information $\{\RI_i\}_{i=1}^n$.
\item The combiner runs the PPT algorithm $\ThrRed(\pk, M, \ADM, \DID, \sigma, \{\RI_i\}^n_{i=1})$.
\paragraph*{${\ThrRed}(\pk, M, \ADM, \DID, \sigma, \{\RI_i\}^n_{i=1}):$}
\begin{enumerate}
    \item Parse $\pk$ as $(\pkADM, \pkM, t, n)$ and $\sigma$ as $(\sigma_{\Fix}, \Sigma_{\agg})$.
    \item Parse $\RI_i$ as $\{\RI_{i,m_j}\}_{m_j \in M}$.    
    \item For $m_j \in M$, define $\RI_{m_j} = \{\RI_{i,m_j}\}_{i=1}^n$. 
    \item Define $\MOD=\{m_j|m_j \in M \land \# \RI_{m_j} \geq t \}$
    \item For $m_j \in {\MOD}$, define $\InRI_{m_j} \leftarrow \{i \in \mathbb{N} | \{\RI_{i,m_j}\} \neq \emptyset \}$.
    \item For $m_j \in {\MOD}$, choose subset $J_{m_j} \subset \InRI_{m_j}$ such that $\# J_{m_j} = t$.
    \item For $m_j \in {\MOD}$, compute $\sigma_{m_j} \leftarrow \prod_{i \in J_{m_j}}\left(\RI_{i, m_j}\right)^{\gamma_{i, J_{m_j}}}$,\\
    where $\gamma_{i, J_{m_j}} = \prod_{j\in J_{m_j}, j \neq i} j(j-i)^{-1}$.
    \item Compute $\sigma_{\MOD} \leftarrow \prod_{{m_j \in \MOD}}\sigma_{m_j}$,\ $\Sigma_{\agg}' \leftarrow \Sigma_{\agg}/ \sigma_{\MOD}$.
    \item Set $M' \leftarrow M \backslash \{\MOD\}$, \ $\sigma' \leftarrow (\sigma_{\Fix}, \Sigma_{\agg}')$.
    \item Return $(M', \ADM, \DID, \sigma')$.
         \end{enumerate}
    \item The combiner outputs $(M', \ADM, \DID, \sigma')$.
\end{enumerate}

\paragraph*{$\Verify(\pk, M, \ADM, \DID, \sigma) :$} Given a tuple $(\pk, M, \ADM, \DID, \sigma)$, the PPT algorithm $\Verify$ works as follows:
\begin{enumerate}
\item Parse $\pk$ as $(\pkADM, \pkM, t, n)$ and $\sigma$ as $(\sigma_{\Fix}, \Sigma_{\agg})$.
\item If $\ADM \cap M \neq \ADM$, return 0.
\item \label{veryhashqadm} Compute $h_{\ADM} \leftarrow H(\DID||\ord({\ADM}))$.\\
$\ord(\ADM)$ denotes a lexicographic ordering to the elements in $\ADM$.
\item \label{veryhashqm} For $m_j \in M$, compute $h_{m_j} \leftarrow H(\DID||m_j)$.\item If $e(\sigma_{\Fix}, g_2) \neq e(h_{\ADM}, \pkADM)$, return 0
\item If $e(\Sigma_{\agg}, g_2) = e \left(h_{\ADM}, \pkM \right) \cdot \prod_{m_j \in M}e(h_{m_j}, \pkM)$, return 1.
Otherwise output 0.
\end{enumerate}

\subsubsection*{Correctness}
If $(M, \ADM, \DID, \sigma)$ is honestly generated by the $\Sign$ and has not been processed by the $\Redact$ protocol, $\Verify(M, \ADM, \DID, \sigma)=1$ always holds.
If  $(M, \ADM, \DID, \sigma)$ is honestly generated the $\Sign$ and $(M', \ADM, \DID, \sigma')$ is honestly redacted from $(M, \ADM, \DID, \sigma)$ by $\Redact$ protocol, $(M', \ADM, \DID, \sigma')$ passes the verification in the $\Verify$. 
Therefore, our construction of $t$-out-of-$n$ redactable signature scheme in the one-time redaction model satisfies correctness.

\subsection{Security of Our $t$-out-of-$n$ Redactable Signature Scheme}

\begin{theorem}\label{ourunf}
In the random oracle model, if the computational co-Diffie-Hellman problem assumption holds, then our proposed $t$-out-of-$n$ redactable signature scheme in the one-time redaction model $\tnRSS$ $\Pi_1$ satisfies the unforgeability property.
\end{theorem}

Here, to explain the outline of the proof, we introduce new notations.
Let $q_s$ be the total number of queries from an adversary to $\mathcal{O}^{\Sign}$, $(M_i, \ADM_i)$ an $i$-th input for $\mathcal{O}^{\Sign}$, $(M^i, \ADM^i, \DID^i, \sigma^i)$ the $i$-th output of $\mathcal{O}^{\Sign}$. We denote
\begin{equation*}
Q_{\Sign}:= \bigcup_{i=1}^{q_s} \{(M^i, \ADM^i, \DID^i)\}, \; \; Q^{\AD}_{\Sign}:= \bigcup_{i=1}^{q_s} \{(\ADM^i, \DID^i)\}. 
\end{equation*}
Also, let $q_r$ be the total number of queries from an adversary to $\mathcal{O}^{\Redact}$,\\ $(M^i, \ADM^i, \DID^i, \sigma^i, \MOD^i)$ an $i$-th input for $\mathcal{O}^{\Redact}$, $(M{'}^{i}, \ADM^{i}, \DID^i, \sigma{'}^{i})$ the $i$-th output of $\mathcal{O}^{\Redact}$. We denote 
\begin{equation*}
Q_{\Redact}:= \bigcup_{i=1}^{q_r} \{(M{'}^{i}, \ADM^{i}, \DID^i)\}, \; \; Q^{\AD}_{\Redact}:= \bigcup_{i=1}^{q_r} \{(\ADM^i, \DID^i)\}.
\end{equation*}

We assume the following three types of PPT adversaries that breaks the unforgeability security in our proposed scheme.
\begin{itemize}
\item An adversary $\A_1$ that outputs a forgery $(M^*, \DID^*, \ADM^*, \sigma^*)$ such that $(\ADM^*, \DID^*) \notin (Q^{\AD}_{\Sign} \cup Q^{\AD}_{\Redact})$.
\item An adversary $\A_2$ that outputs a forgery $(M^*, \DID^*, \ADM^*, \sigma^*)$ which satisfies $(\ADM^*, \DID^*) \in (Q^{\AD}_{\Sign} \cup Q^{\AD}_{\Redact})$. Moreover, there is $\tilde{M}$ such that $(\tilde{M}, \ADM^*, \DID^*) \in (Q_{\Sign} \cup Q_{\Redact})$.
\item An adversary $\A_3$ that outputs a forgery $(M^*, \DID^*, \ADM^*, \sigma^*)$ which satisfies $(\ADM^*, \DID^*) \in (Q^{\AD}_{\Sign} \cup Q^{\AD}_{\Redact})$. Moreover, there are no $\tilde{M}$ such that $(\tilde{M}, \ADM^*,$ $\DID^*) \in (Q_{\Sign} \cup Q_{\Redact})$ and $\tilde{M}\nsubseteq M$.
\end{itemize}

To prove the theorem, for each $\A_i$, we consider a sequential of games from the original unforgeability game to game which is directly related to solving a $\coCDH$ problem. Then, We construct $\B_i$ which breaking the $\coCDH$ assumption by running $\A_i$.
$\B_1$ breaks the $\coCDH$ assumption by using the forgery $\sigma^*_{\Fix}$. In the case of $\B_2$ and $\B_3$, they use the forgery $\Sigma_{\agg}^*$ to break the $\coCDH$ assumption. 
One difference between $\B_2$ and $\B_3$ is how to program the hash value.
\subsubsection*{Case $1$}
We consider an adversary $\A_1$ that can generate a valid forgery with $\epsilon_{\uf1}$ against our proposal redactable signature scheme.
Let $\game_{1-0}$ be the original unforgeability game in a redactable signature scheme and $\game_{1-5}$ be directly related to solving the computational co-Diffie-Hellman problem. 
Define $\Adv_{\A_1}[\game_{1-X}]$ as the advantage of an adversary $\mathcal{\A}_1$ in $\game _{1-X}$.
\begin{itemize}
 \item $\game_{1-0}$: Original unforgeability game in a redactable signature scheme.  
 \begin{equation*}
  \Adv_{\A_1}[\game_{1-0}] =\epsilon_{{\sf \uf1}}
 \end{equation*}
 \item $\game_{1-1}$: We change a key generation algorithm $\KeyGen$ in Step \ref{keygenfix}.\\
 Choose $\tilde{x} \xleftarrow{\$} \mathbb{Z}_q,\; \tilde{r} \xleftarrow{\$} \mathbb{Z}_q$ and compute $u \leftarrow g^{\tilde{x}}$, $\tilde{y} \leftarrow g_2^{\tilde{x}+\tilde{r}}$. \\
Set $(\pkADM, \skADM) \leftarrow (\tilde{y}, \tilde{x}+\tilde{r})$.
 \item $\game_{1-2}$:We change a setting of the random oracle $\mathcal{O}^H$. 
 Fix $h \xleftarrow{\$} \mathbb{G}_2$ and let $\mathbb{T}$ be a table that maintains a list of tuples $\langle v, w, b, c \rangle$ as explain below.
 We refer to this list for the query to $\mathcal{O}^h$. The initial state of $\mathbb{T}$ is empty.
 For queries $v^{(i)}$ to $\mathcal{O}^H$:
 \begin{itemize}
 \item If $\langle v^{(i)}, w^{(i)}, \cdot, \cdot \rangle$ (Here, `$\cdot$' represents an arbitrary value) already appears in $\mathbb{T}$, then return $w^{(i)}$.
 \item Choose $s^{(i)} \xleftarrow{\$} \mathbb{Z}_q$.
 \item Flip a biased coin $c^{(i)} \in \{0, 1\}$ such that $\Pr[c^{(i)}=0] = 1-1/(q_s + 1)$ and $\Pr[c^{(i)}=1] = 1/(q_s + 1)$.
 \item If $c^{(i)}=0$, compute $w^{(i)}=\phi(g_2)^{b^{(i)}}$. 
 \item If $c^{(i)}=1$, compute $w^{(i)}=h\cdot \phi(g_2)^{b^{(i)}}$.
 \item Insert $\langle v^{(i)}, w^{(i)}, s^{(i)}, c^{(i)} \rangle$ in $\mathbb{T}$ and return $w^{(i)}$.
 \end{itemize}
 \item $\game_{1-3}$: We modify the signing algorithm $\Sign$ in Step \ref{signhashqadm} as follows:
 \begin{itemize}
 \item Set $v^{(0)} \leftarrow (\DID||\ord(\ADM))$.
 \item Query $v^{(0)}$ to $\mathcal{O}^H$. We assume $\langle v^{(0)}, w^{(0)}, b^{(0)}, c^{(0)} \rangle$ to be the tuple in $\mathbb{T}$ for $v^{(0)}$.
 \item If $c^{(0)}=1$, return $\perp$ and abort.
 \end{itemize}
 \item $\game_{1-4}$: We modify the signing algorithm $\Sign$ in Step \ref{signfix} as follows:
 \begin{itemize}
  \item Compute $\sigma_{\Fix} \leftarrow \phi(u)^{b^{(0)}}\cdot \phi(g_2)^{\tilde{r}{b^{(0)}}}$.
 \end{itemize}
 (A signature $\sigma_{\Fix}$ can be generated without a knowledge of $\skADM$.)
 
 \item $\game_{1-5}$: We receive a valid forgery $(M^*, \ADM^* \DID^*, \sigma^*)$ from the adversary $\A_1$, we operate as follows:
 \begin{itemize}
 \item Set $v^{(0)} \leftarrow (\DID^*||\ord(\ADM^*))$.
 \item Query $v^{(0)}$ to $\mathcal{O}^H$. We assume $\langle v^{(0)}, w^{(0)}, s^{(0)}, c^{(0)} \rangle$ to be the tuple in $\mathbb{T}$ for each $v^{(0)}$.
 \item If $c^{(0)} = 0$, then abort. 
 \end{itemize}
 \end{itemize} 

\begin{lemma}\label{A1sta} The following equation holds.
\begin{equation*}
\Adv_{\A_1}[\game_{1-1}] = \Adv_{\A_1}[\game_{1-0}].
\end{equation*}
\end{lemma}
Since the distribution of $(\pk_{\Fix}, \sk_{\Fix})$ in $\game_{1-0}$ and $\game_{1-1}$ are same. 

\begin{lemma}
If $H$ is the random oracle model, the following eqauation holds.
\begin{equation*}
\Adv_{\A_1}[\game_{1-2}] = \Adv_{\A_1}[\game_{1-1}]
\end{equation*}
\end{lemma}
 Since the distribution of outputs of $\mathcal{O}^H$ in $\game_{1-1}$ and $\game_{1-2}$ are identical.

\begin{lemma} The following inequality holds.
\begin{equation*}
\Adv_{\A_1}[\game_{1-3}] \geq (1-1/(q_s+1))^{q_s} \times \Adv_{\A_1}[\game_{1-2}].
\end{equation*}
\end{lemma}
Since the probability that each signing query does not abort at least $1-1/(q_s+1)$.

\begin{lemma} The following equation holds.
\begin{equation*}
\Adv_{\A_1}[\game_{1-4}] = \Adv_{\A_1}[\game_{1-3}].
\end{equation*}
\end{lemma}
Since outputs of $\Sign$ in $\game_{1-3}$ and $\game_{1-4}$ are same.

\begin{lemma}\label{A1gol} The following inequality holds.
\begin{equation*}
\Adv_{\A_1}[\game_{1-5}] \geq (1/(q_s +1)) \times \Adv_{\A_1}[\game_{1-4}].
\end{equation*}
\end{lemma}
Since the probability that the forged signature satisfies $c^{(0)}=1$ at least $1/(q_s+1)$.

To summarize {\bf Lemma \ref{A1sta}} to {\bf \ref{A1gol}}, the following holds.\\
(In the following equation, $e$ represents the Napier's constant.)
\begin{equation*}
\begin{split}
\Adv_{\A_1}[\game_{1-5}] &\geq  (1-1/(q_s+1))^{q_s}\times (1/(q_s +1)) \times \Adv_{\A_1}[\game_{1-0}]\\
& \geq (1/e) \times (1/(q_s+1)) \times \Adv_{\A_1}[\game_{1-0}]
\end{split}
\end{equation*}
 
Now we construct the algorithm $\B_1$ which breaking the computational co-Diffie-Hellman assumption by running the algorithm $\A_1$.
The operation of $\B_1$ for the input co-Diffie-Hellman problem instance $(g_2, g_2^{\alpha}, h^*)$ is changed to $h$ to $h^*$ and $u$ to $g_2^{\alpha}$ in $\game_{1-5}$. 
Suppose $\B_1$ does not abort receiving a forgery $(M^*, \ADM^*, \DID^*, \sigma^*)$ from $\A_1$.\\
$\B_1$ parses $\sigma^*$ as $(\sigma^*_{\Fix^*}, \Sigma_{\agg}^*)$, sets $v^{(0)} \leftarrow (\DID^*||\ord(\ADM^*))$
and computes $w^{(0)} \leftarrow h^* \cdot \phi(g_2)^{b^{(0)}}$. 
Since $(M^*, \ADM^*, \DID^*, \sigma^*)$ is valid and $\pkADM = g_2^{{\alpha}+\tilde{r}}$,
$e(\sigma^*_{\Fix}, g_2) = e((w^{(0)})^{{\alpha}+\tilde{r}}, g_2)$ holds. 
It implies that $\sigma^*_{\Fix} = (w^{(0)})^{\alpha+\tilde{r}}=(h^* \cdot \phi(g_2)^{b^{(0)}})^{\alpha+\tilde{r}}$.
Therefore, $\B_1$ computes $(h^*)^{\alpha} = \sigma_{\Fix}^* \cdot (\phi(u)^{b^{(0)}} \cdot (h^*)^{\tilde{r}} \cdot \phi(g_2)^{\tilde{r}b^{(0)}})^{-1}$ and outputs the solution $(h^*)^{\alpha}$ of the computational co-Diffie-Hellman problem instance $(g_2, g_2^{\alpha}, h^*)$.

Let $\epsilon_{{\sf co\mathchar`-cdh}}$ is the probability that $\B_1$ breaks the computational co-Diffie-Hellman assumption.
We can bound the probability $\epsilon_{{\sf co\mathchar`-cdh1}} \geq \Adv_{\A_1}[{\game}\ 1-5]$ and $\epsilon_{{\sf co\mathchar`-cdh1}} \geq (1/e) \times (1/q_s+1) \times \epsilon_{{\sf uf1}}$ holds. ($e$ represents the Napier's constant.)
Hence, if $\epsilon_{{\sf uf1}}$ is non-negligiable in $\lambda$, $\B_1$ breaks the computational co-Diffie-Hellman assumption with non-negligiable in $\epsilon_{{\sf co\mathchar`-cdh1}}$. 

\subsubsection*{Case $2$}
We consider an adversary $\A_2$ that can generate a valid forgery with $\epsilon_{{\sf uf2}}$ against our proposal redactable signature scheme.
Let $\game_{2-0}$ be the original unforgeability game in a redactable signature scheme and $\game_{2-6}$ be directly related to solve the computational co-Diffie-Hellman problem. 
Define $\Adv_{\A_2}[\game_{2-X}]$ as the advantage of an adversary $\mathcal{\A}_2$ in $\game_{2-X}$.
\begin{itemize}
 \item $\game_{2-0}$: Original unforgeability game in a redactable signature scheme.  
 \begin{equation*}
  \Adv_{\A_2}[\game_{2-0}] =\epsilon_{{\sf uf2}}
\end{equation*}
 \item $\game_{2-1}$: We change a setting of $\mathcal{O}^{\Redact}$.\\
 We introduce a table $\mathbb{L}^u$ that store $\DID$s and $\mathbb{L}^{0}=\emptyset$.\\
 For a $u$-th query $(M, \ADM, \DID, \sigma, \MOD)$ to $\mathcal{O}^{\Redact}$:
 \begin{itemize}
 \item Parse $\pk$ as $(\pkADM, \pkM, t, n)$ and $\sigma$ as $(\sigma_{\Fix}, \Sigma_{\agg})$.
 \item If $\DID \in \mathbb{L}^{u-1}$, then abort. 
 \item Set $\mathbb{L}^{u} \leftarrow \mathbb{L}^{u-1} \cup \{\DID\}$.
 \item If $\MOD \nsubseteq M \lor \MOD \cap \ADM \neq \emptyset$, then abort.
 \item Compute $h_{\ADM} \leftarrow H(\DID||\ord({\ADM}))$.\\
$\ord(\ADM)$ denotes a lexicographic ordering to the elements in $\ADM$.
 \item For $m_j \in M$, compute $h_{m_j} \leftarrow H(\DID||m_j)$.\item If $e(\sigma_{\Fix}, g_2) \neq e(h_{\ADM}, \pkADM)$, then abort.
 \item If $e(\Sigma_{\agg}, g_2) \neq e \left(h_{\ADM}, \pkM \right) \cdot \prod_{m_j \in M}e(h_{m_j}, \pkM)$, then abort.
 \item For $m_j \in \MOD$, compute $\sigma_{m_j} \leftarrow H(\DID||m_j)^{\skM}$.
 \item Compute $\sigma_{\MOD} \leftarrow \prod_{m_j \in \MOD} \sigma_{m_j}$, $\Sigma_{\agg}' \leftarrow \Sigma_{\agg} / \sigma_{\MOD}$.
 \item Set $M' \leftarrow M \backslash \MOD$, $\sigma' \leftarrow (\sigma_{\Fix}, \Sigma_{\agg}')$.
 \item Return $(M', \ADM, \DID, \sigma')$.
 \end{itemize}
 (Redactions are done using $\skM$ instead of using $\{\rk[i]\}^n_{i=1}$.)
 
 \item $\game_{2-2}$: We change settings of $\KeyGen$ and $\mathcal{O}^{\Corrupt}$.
 \begin{itemize}
 \item We change a key generation algorithm $\KeyGen$ in Step \ref{keygenm} to \ref{keygenout}. 
 \begin{itemize}
 \item Choose $x \xleftarrow{\$} \mathbb{Z}_q,\; r \xleftarrow{\$} \mathbb{Z}_q$, compute $u \leftarrow g^x$, $y \leftarrow g_2^{x+r}$.
 \item Set $\pkM \leftarrow y, \; \skM \leftarrow x + r$.
 \item Return $(\pk, \sk) \leftarrow ((\pkADM, \pkM, t, n), (\skADM, \skM))$.
 \end{itemize}
(Redactor's keys $\{\rk[i]\}^n_{i=1}$ are not generated in the $\KeyGen$.)
 \item We change the setting of $\mathcal{O}^{\Corrupt}$ as follows:\\
 Let $CR$ is a list to store a redactor's key information $(i, \rk[i])$\\
 For a query $i$ to $\mathcal{O}^{\Corrupt}$,
 \begin{itemize}
 \item If $(i, \rk[i])$ already appears in $CR$, then return $\rk[i]$.
 \item Choose $f(i) \xleftarrow{\$} \Z_q$, set $CR \leftarrow CR \cup \{(i, f(i))\}$.
 \item Return $\rk[i] \leftarrow (i, f(i))$.
 \end{itemize}
 \end{itemize}

 \item $\game_{2-3}$: We change a setting of the random oracle $\mathcal{O}^H$. 
 Fix $h \xleftarrow{\$} \G_2$ and let $\mathbb{T}$ be a table that maintains a list of tuples $\langle v, w, b, c \rangle$ as explain below.
 We refer to this list for the query to $\mathcal{O}^h$. The initial state of $\mathbb{T}$ is empty.
 For queries $v^{(i)}$ to $\mathcal{O}^H$:
 \begin{itemize}
 \item If $\langle v^{(i)}, w^{(i)}, \cdot, \cdot \rangle$ (Here, `$\cdot$' represents an arbitrary value) already appears in $\mathbb{T}$, then return $w^{(i)}$.
 \item Choose $s^{(i)} \xleftarrow{\$} \mathbb{Z}_q$.
 \item Flip a biased coin $c^{(i)} \in \{0, 1, 2\}$  such that such that $\Pr[c^{(i)}=1] = 1-1/((\ell+1)(q_s+q_r)+1)$, $\Pr[c^{(i)}=1] = 1/(2(\ell+1)(q_s+q_r)+2)$, $\Pr[c^{(i)}=2] = 1/(2(\ell+1)(q_s+q_r)+2)$.
 \item If $c^{(i)}=0$, compute $w^{(i)}=\phi(g_2)^{b^{(i)}}$. 
 \item If $c^{(i)}=1$, compute $w^{(i)}=h\cdot \phi(g_2)^{b^{(i)}}$.
 \item If $c^{(i)}=2$, compute $w^{(i)}=h^{-1}\cdot \phi(g_2)^{b^{(i)}}$.
 \item Insert $\langle v^{(i)}, w^{(i)}, s^{(i)}, c^{(i)} \rangle$ in $\mathbb{T}$ and return $w^{(i)}$.
 \end{itemize}
 
 \item $\game_{2-4}$:We modify the signing algorithm $\Sign$ in Step 6 as follows:
 \begin{itemize}
 \item Set $v^{(0)} \leftarrow (\DID||\ord(\ADM))$, $v^{(j)} \leftarrow (\DID||m_j) \; (1\leq j \leq \#M)$.
 \item Query $v^{(j)}$ $(0 \leq j \leq \#M)$ to $\mathcal{O}^H$. We assume $\langle v^{(j)}, w^{(j)}, b^{(j)}, c^{(j)} \rangle$ to be the tuple in $\mathbb{T}$ for each $v^{(j)}$ $(1 \leq j \leq \#M)$.
 \item If $c^{(0)}=2$, $c^{(1)}=1$, $c^{(j)}=0$ $(2 \leq \forall j \leq \#M)$ or $c^{(j)}=0$ $(0 \leq \forall j \leq \#M)$, go to Step \ref{signfix} of $\Sign$. Otherwise return $\perp$ and abort.
 \end{itemize}
 
  \item $\game_{2-5}$: We modify the signing algorithm $\Sign$ in Step \ref{signm}, \ref{signagg} as follows:
 \begin{itemize}
 \item If $c^{(0)}=2$, $c^{(1)}=1$, $c^{(j)}=0$ $(2 \leq \forall j \leq \#M)$,
  \begin{itemize}
  \item Compute $\sigma_{\ADM m_1} \leftarrow  \phi(u)^{b^{(0)}+b^{(1)}}\cdot \phi(g_2)^{r(b^{(0)}+b^{(1)})}$.
  \item For all $m_j \in M \backslash \{m_1\}$, compute $\sigma_{m_j} \leftarrow \phi(u)^{b^{(j)}}\cdot \phi(g_2)^{r{b^{(j)}}}$.
  \item Compute $\Sigma_{\agg} \leftarrow \sigma_{\ADM m_1} \cdot \prod_{m_j \in M \backslash \{m_1\}} \sigma_{m_j}$.
  \end{itemize}
 \item If $c^{(j)}=0$ $(0 \leq \forall j \leq \#M)$,
  \begin{itemize} 
  \item Compute $\sigma_{\ADM} \leftarrow \phi(u)^{b^{(0)}}\cdot \phi(g_2)^{r{b^{(0)}}}$.
  \item For all $m_j \in M$, compute $\sigma_{m_j} \leftarrow \phi(u)^{b^{(j)}}\cdot \phi(g_2)^{r{b^{(j)}}}$.
  \item Compute $\Sigma_{\agg} \leftarrow \sigma_{\ADM} \cdot \prod_{m_j \in M} \sigma_{m_j}$.
  \end{itemize}
 \end{itemize}
 
 (By above modification, a signature $\Sigma_{\agg}$ can be generated without a knowledge of the $\skM$.)

 \item $\game_{2-6}$: We change a setting of $\mathcal{O}^{\Redact}$.
 \begin{itemize}
 \item Parse $\pk$ as $(\pkADM, \pkM, t, n)$ and $\sigma$ as $(\sigma_{\Fix}, \Sigma_{\agg})$.
 \item If $\DID \in \mathbb{L}^{u-1}$, then abort. 
 \item Set $\mathbb{L}^{u} \leftarrow \mathbb{L}^{u-1} \cup \{\DID\}$.
 \item If $\MOD \nsubseteq M \lor \MOD \cap \ADM \neq \emptyset$, then abort.
 \item Set $v^{(0)} \leftarrow (\DID||\ord(\ADM))$, $v^{(j)} \leftarrow (\DID||m_j) \; (1\leq j \leq \#M)$.
 \item Query $v^{(j)}$ $(0 \leq j \leq \#\MOD)$ to $\mathcal{O}^H$. 
 We assume $\langle v^{(j)}, w^{(j)}, b^{(j)}, c^{(j)} \rangle$ to be the tuple in $\mathbb{T}$ for each $v^{(j)}$ $(1 \leq j \leq \#\MOD)$.
 \item If $e(\sigma_{\Fix}, g_2) \neq e(w^{(0)}, \pkADM)$, then abort.
 \item If $e(\Sigma_{\agg}, g_2) \neq \prod_{0 \leq j \leq \#M}e(w^{(j)}, \pkM)$, then abort.

 \item If $c^{(j)}=0$ $(\forall m_j \in \MOD)$, go to next step. Otherwise return $\perp$ and abort.
 \item For all $m_j \in \MOD$, compute $\sigma_{m_j} \leftarrow \phi(u)^{b^{(j)}}\cdot \phi(g_2)^{r{b^{(j)}}}$.
 \item Compute $\sigma_{\MOD} \leftarrow \prod_{m_j \in \MOD} \sigma_{m_j}$, $\Sigma_{\agg}' \leftarrow \Sigma_{\agg} / \sigma_{\MOD}$.
 \item Set $M' \leftarrow M \backslash \MOD$, $\sigma' \leftarrow (\sigma_{\Fix}, \Sigma_{\agg}')$.
 \item Return $(M', \ADM, \DID, \sigma')$.
 \end{itemize}
 (Redactions can be done without the knowledge of the $\skM$.)

 \item $\game_{2-7}$: We receiving the output forgery $(M^*, \ADM^*, \DID^*, \sigma^*)$ from the adversary $\A_3$, 
 \begin{itemize}
 \item Set $v^{(0)} \leftarrow (\DID^*||\ord(\ADM^*))$, $v^{(j)} \leftarrow (\DID||m_j^*) \; (1\leq j \leq \#M^*)$.
 \item Query $v^{(j)}$ $(0 \leq j \leq \#M^*)$ to $\mathcal{O}^H$. We assume $\langle v^{(j)}, w^{(j)}, s^{(j)}, c^{(j)} \rangle$ to be the tuple in $\mathbb{T}$ for each $v^{(j)}$ $(0 \leq j \leq \#M^*)$.
 \item If $c^{(0)} = 1$ and $c^{(j)} = 0$ $(1 \leq j \leq \#M^*)$, then accept. Otherwise reject and abort. 
 \end{itemize}

\begin{lemma}\label{A2sta} The following equation holds. 
\begin{equation*}
\Adv_{\A_2}[\game_{2-1}] = \Adv_{\A_2}[\game_{2-0}].
\end{equation*}
\end{lemma}
Since outputs of $\mathcal{O}^{\Redact}$ in $\game_{2-0}$ and $\game_{2-1}$ are same.

\begin{lemma}\label{A2nex} The following equation holds.
\begin{equation*}
\Adv_{\A_2}[\game_{2-2}] = \Adv_{\A_2}[\game_{2-1}].
\end{equation*}
\end{lemma}
To simplify the discussion,  let $\A_2$ get $\rk[i],\dots, \rk[t-1]$ from $\mathcal{O}^{\Corrupt}$.
In $[\game_{2-1}]$, the following equation holds.
\begin{equation*}
V
  \left(
   \begin{array}{c}
   a_0\\
   a_1\\
   a_2\\
   \vdots \\
   a_{t-1}\\
    \end{array}
  \right)
  =
  \left(
   \begin{array}{c}
   f(0)\\
   f(1)\\
   f(2)\\
   \vdots \\
   f(t-1)\\
    \end{array}
  \right)
    \; {\rm where} \;
  V=
     \left(
    \begin{array}{ccccc}
      1 & 0 & 0 & \cdots & 0 \\
      1 & 1 & 1 & \cdots & 1\\
      1 & 2 &  2^2& \cdots & 2^{t-1}\\
       \vdots & \vdots & \vdots & \cdots & \vdots\\
      1& t-1& (t-1)^2&\cdots & (t-1)^{t-1}
    \end{array}
  \right).
\end{equation*}
Since $V$ is the Vandermonde matrix, $V$ is the regular matrix. Distributions of $(a_0, a_1, \cdots, a_{t-1})$ and $(f(0), f(1), \dots, f(t-1))$ are identical.
Therefore, distributions of $(\skM, \rk[1], \dots, \rk[t-1])$ in $[\game_{2-1}]$ and $[\game_{2-2}]$ are same.

\begin{lemma}
If $H$ is the random oracle model, the following equation holds.
\begin{equation*}
\Adv_{\A_2}[\game_{2-3}] = \Adv_{\A_2}[\game_{2-2}].
\end{equation*}
\end{lemma}
Since the distribution of outputs of $\mathcal{O}^H$ in $\game_{2-3}$ and $\game_{2-2}$ is identical.

\begin{lemma} The following inequality holds.
\begin{equation*}
\Adv_{\A_2}[\game_{2-4}] \geq (1-1/((\ell +1 )(q_s+q_r)+1))^{(\ell+1)q_s} \times \Adv_{\A_2}[\game_{2-3}].
\end{equation*}
\end{lemma}
Since the probability that each signing query does not abort at least \\$(1-1/((\ell +1 )(q_s+q_r)+1))^{(\ell+1)}$.

\begin{lemma} The following equation holds.
\begin{equation*}
\Adv_{\A_2}[\game_{2-5}] = \Adv_{\A_2}[\game_{2-4}].
\end{equation*}
\end{lemma}
Since outputs of $\Sign$ in $\game_{2-5}$ and $\game_{2-4}$ are same.

\begin{lemma} The following inequality holds.
\begin{equation*}
\Adv_{\A_2}[\game_{2-6}] \geq (1-1/((\ell +1 )(q_s+q_r)+1))^{(\ell+1)q_r} \times \Adv_{\A_2}[\game_{2-5}].
\end{equation*}
\end{lemma}
Since the probability that each redaction query does not abort at least \\$(1-1/((\ell +1 )(q_s+q_r)+1))^{(\ell+1)}$.

\begin{lemma}\label{A2gol} The following inequality holds.
\begin{equation*}
\begin{split}
\Adv_{\A_2}&[\game_{2-7}] \\
&\geq \frac{\left(\frac{1}{2(\ell+1)(q_s+q_r)+2} \right)^2}{\left(1-\frac{1}{(\ell +1 )(q_s+q_r)+1}\right)^2 +  \left(\frac{1}{2(\ell+1)(q_s+q_r)+2} \right)^2} \times \Adv_{\A_2}[\game_{2-6}]\\
&= (1/(4(\ell+1)^2(q_s+q_r)^2 + 1))\times \Adv_{\A_2}[\game_{2-6}].
\end{split}
\end{equation*}
\end{lemma}
Since an output $(M^*, \ADM^*, \DID^*, \sigma^*)$ satisfies $(c^{(0)}, c^{(1)}) = (0,0)$ or $(2, 1)$.

\end{itemize}
To summarize {\bf Lemma \ref{A2sta}} to {\bf \ref{A2gol}}, the following holds.\\
(In the following equation, $e$ represents the Napier's constant.)
\begin{equation*}
\begin{split}
\Adv_{\A_2}[\game_{2-7}] &\geq  (1-1/((\ell +1 )(q_s+q_r)+1))^{(\ell+1)(q_s+q_r)}\\
&\;\;\;\;\;\times 1/(4(\ell+1)^2(q_s+q_r)^2 + 1) \times \Adv_{\A_2}[\game_{2-0}]\\
& \geq (1/e) \times (1/(4(\ell+1)^2(q_s+q_r)^2 + 1)) \times \Adv_{\A_2}[\game_{2-0}]
\end{split}
\end{equation*}

Now we construct the algorithm $\B_2$ which breaking the computational co-Diffie-Hellman assumption by running the algorithm $\A_2$.
The operation of $\B_2$ for the input co-Diffie-Hellman problem instance $(g_2, g_2^{\alpha}, h^*)$ is changed to $h$ in $\game_{2-7}$ to $h^*$ and $u$ to $g_2^{\alpha}$.\\ 
Suppose $\B_2$ do not abort receiving a forgery $(M^*, \ADM^*, \DID^*, \sigma^*)$ from $\A_2$.
$\B_3$ parses $\sigma^*$ as $(\sigma_{\ADM^*}^*, \Sigma_{\agg}^*)$, sets $v^{(j)} \leftarrow ({\sf DID}^*||m_j^*) \; (1\leq j \leq \#M^*)$, and computes $w^{(1)} \leftarrow h\cdot \phi(u)^{b^{(1)}}\cdot \phi(g_2)^{r{b^{(1)}}}$,  $w^{(j)} \leftarrow \phi(u)^{b^{(j)}}\cdot \phi(g_2)^{r{b^{(j)}}} \;(2 \leq j \leq \#M^*)$.
Then $\B_3$ computes $\sigma^*_{m^*_1} \leftarrow \Sigma_{\agg}^*/ \prod^{\#M*}_{j=2} \sigma_{m_j}$.
Since $(M^*, \ADM^*, \DID^*, \sigma^*)$ is valid signature and $\pkM = g_2^{{\alpha}+r}$,
$e(\sigma^*_{m^*_1}, g_2) = e\left((w^{(1)})^{\alpha+r}, g_2\right)$ holds.
It implies that $\sigma^*_{m^*_1} = (w^{(1)})^{\alpha+r}=(h^* \cdot \phi(g_2)^{b^{(1)}})^{\alpha+r}$.
Therefore, $\B_3$ computes $(h^*)^{\alpha} = \sigma^*_{m^*_1} \cdot (\phi(u)^{b^{(1)}} \cdot (h^*)^r \cdot \phi(g_2)^{rb^{(1)}})^{-1}$ and outputs the solution $(h^*)^{\alpha}$ of the computational co-Diffie-Hellman problem instance $(g_2, g_2^{\alpha}, h^*)$.

Let $\epsilon_{{\sf co\mathchar`-cdh2}}$ is the probability that $\B_2$ breaks the computational co-Diffie-Hellman assumption.
We can bound the probability $\epsilon_{{\sf co\mathchar`-cdh2}} \geq \Adv_{\A_2}[\game_{2-7}]$ and $\epsilon_{{\sf co\mathchar`-cdh2}} \geq  (1/e) \times (1/(4(\ell+1)^2(q_s+q_r)^2 + 1)) \times \epsilon_{{\sf uf2}}$ holds. ($e$ represents the Napier's constant.)
If $\epsilon_{{\sf uf2}}$ is non-negligiable in $\lambda$, $\B_2$ breaks the computational co-Diffie-Hellman assumption with non-negligiable in $\epsilon_{{\sf co\mathchar`-cdh2}}$. 
We will describe the proof of {\bf Case 3} in the full version of this paper. \qed

\begin{theorem}\label{ourtrans}
Our proposed $t$-out-of-$n$ redactable signature scheme in the one-time redaction model $\tnRSS$ $\Pi_1$ satisfies the transparency.
\end{theorem}
We will describe the proof of {\bf Theorem \ref{ourtrans}} in the full version of this paper.
By {\bf Theorem \ref{u-tra-pri}} and  {\bf Theorem \ref{ourtrans}}, our proposed scheme satisfies the privacy.

\section{Conclusion}\label{concludepaper}
In this paper, we introduce the new notion of $t$-out-of-$n$ $\RSS$. Our proposed model supports only the one-time redaction model which allows redacting signed message only one time for each signature. 
Our construction $\Pi_1$ does not satisfy the unforgeability in a model that allows redacting signed message many times.
For example, $M=\{m_1, m_2, m_3\}$ and $\ADM=\emptyset$, an adversary who does the following operation generates a valid forgery in a multiple redactions model.

\begin{enumerate}
\item Given $\pk$ from $\C$.
\item Query $(M, \ADM)$ to $\mathcal{O}^{\Sign}$ and get $(M, \ADM, \DID, \sigma)$.
\item Let $\MOD^1 = \{m_1\}$ $\MOD^2 = \{m_2\}$. Query $(M, \ADM, \DID, \sigma, \MOD^1)$ to $\mathcal{O}^{\Redact}$ and get $(M', \ADM, \DID, \sigma')$ and query $(M', \ADM, \DID, \sigma', \MOD^2)$ to $\mathcal{O}^{\Redact}$ and get $(M'', \ADM, \DID, \sigma'')$.
\item Parse $\sigma$ as $(\sigma_{\Fix}, \Sigma_{\agg})$, $\sigma'$ as $(\sigma'_{\Fix}, \Sigma_{\agg}')$, and $\sigma''$ as $(\sigma''_{\Fix}, \Sigma_{\agg}'')$.
\item Compute $\sigma_{m_1} \leftarrow \Sigma_{\agg} \cdot (\Sigma_{\agg}')^{-1}$, $\Sigma_{\agg}^* \leftarrow \sigma_{m_1} \cdot \Sigma_{\agg}''$.
\item Set $M^* \leftarrow \{m_1, m_3\}$, $\sigma^* \leftarrow (\sigma_{\Fix}, \Sigma_{\agg}^*)$ and output $(M^*, \DID, \ADM, \sigma^*)$
\end{enumerate}
Giving a construction of $\RSS$ in the multiple redactions model is the interesting futures work.
\section*{Acknowledgement}
A part of this work was supported by 
NTT Secure Platform Laboratories,
JST OPERA JPMJOP1612, JST CREST JPMJCR14D6, and JSPS KAKENHI JP16H01705, JP17H01695.
We would like to thank anonymous referees for their constructive comments.
Based on their comments, we would be more than happy to improve this paper.

\bibliographystyle{abbrvurl}
\bibliography{refR}

%
\appendix

\section{Proof of Theorem \ref{u-tra-pri}} \label{u-tra-pripoof}

\begin{proof}
Assume that PPT adversary $\A^{\priv}$ that wins the privacy game with probability $1/2 + \epsilon_{\priv}$ where $\epsilon_{\priv}$ is non-negligible in $\lambda$.
Let $\C^{\tran}$ be the challenger in transparency game.
Now we construct a PPT adversary $\B^{\tran}$ that wins the transparency game with probability $1/2 + \epsilon_{\priv}/2$ running $\A^{\priv}$. The operation of $\B^{\tran}$ is following.
\begin{itemize}
\item $\B^{\tran}$ receives $\pk$ from $\C^{\tran}$, chooses a bit $c \leftarrow \{0, 1\}$ and sends $\pk$ to $\A^{\priv}$.
\item For each query $(M, \ADM)$ of $\A^{\priv}$ to $\mathcal{O}^{\Sign}$, $\B^{\tran}$ queries $(M, \ADM)$ to $\mathcal{O}^{\Sign}$ and gets $(M, \ADM, \DID, \sigma)$ and sends it to $\A^{priv}$.
\item For each query $(M^0, \ADM^0, \MOD^{0}, M^1, \ADM^1, \MOD^{1})$ of $\A^{\priv}$ to $\mathcal{O}^{\LR}$, $\B^{\tran}$ checks $M^0{'} = M^1{'}$ where $M^0{'} = M^0 /\MOD^0$ and $M^1{'} = M^1/\MOD^1$.
If so, $\B^{\tran}$ queries $(M^c, \ADM^c, \allowbreak \MOD^c)$ to $\mathcal{O}^{\SR}$ and $\B^{\tran}$ returns its result to $\A^{\priv}$.
Otherwise, $\B^{\tran}$ returns $\bot$ to~$\A^{\priv}$.
\item $\B^{\tran}$ receives a guess $b^*$ from $\A^{\priv}$.
If $b^*=c$, $\B^{\tran}$ outputs 0, otherwise $\B^{\tran}$ outputs 1.
\end{itemize}
If $b=0$, $\mathcal{O}^{\SR}$ always redacts and the view of $\A^{\priv}$ is the same as in the privacy game.
However, if $b=1$, each signature is fresh and the output of $\A^{\priv}$ is useless to win the transparency game.
Therefore, the win probability of $\B^{\tran}$ in transparency game is $\epsilon_{\tran} = 1/2(1/2 + \epsilon_{\priv}) +1/2 \cdot 1/2
 = 1/2 + \epsilon_{\priv}/2$.
Therefore, the advantage of $\B^{\tran}$ in transparency game is non-negligible in $\lambda$.  \qed
\end{proof}

\section{Proof of Case 3 in Theorem \ref{ourunf}} \label{unfpoof}
\begin{proof}
We consider an adversary $\A_3$ that can generate a valid forgery with $\epsilon_{{\sf uf3}}$ against our proposal redactable signature scheme.
Let $\game_{3-0}$ be the original unforgeability game in a redactable signature scheme and $\game_{3-6}$ be directly related to solve the computational co-Diffie-Hellman problem. 
Define $\Adv_{\A_3}[\game_{3-X}]$ as the advantage of an adversary $\mathcal{\A}_3$ in $\game_{3-X}$.
\begin{itemize}
 \item $\game_{3-0}$: Original unforgeability game in a redactable signature scheme.  
 \begin{equation*}
  \Adv_{\A_3}[\game_{3-0}] =\epsilon_{{\sf uf3}}
\end{equation*}
 \item $\game_{3-1}$: $\game_{3-1}$ is the same as $\game_{2-1}$.
 \item $\game_{3-2}$: $\game_{3-2}$ is the same as $\game_{2-2}$.
 \item $\game_{3-3}$: We change a setting of the random oracle $\mathcal{O}^H$. 
 Fix $h \xleftarrow{\$} \G_2$ and let $\mathbb{T}$ be a table that maintains a list of tuples $\langle v, w, b, c \rangle$ as explain below.
 We refer to this list for the query to $\mathcal{O}^h$. The initial state of $\mathbb{T}$ is empty.
 For queries $v^{(i)}$ to $\mathcal{O}^H$:
 \begin{itemize}
 \item If $\langle v^{(i)}, w^{(i)}, \cdot, \cdot \rangle$ (Here, `$\cdot$' represents an arbitrary value) already appears in $\mathbb{T}$, then return~$w^{(i)}$.
 \item Choose $s^{(i)} \xleftarrow{\$} \mathbb{Z}_q$.
 \item Flip a biased coin $c^{(i)} \in \{0, 1\}$ such that  $\Pr[c^{(i)}=0] = 1- 1/((\ell+1)(q_s+q_r)+\ell)$,  $\Pr[c^{(i)}=1] = 1/((\ell+1)(q_s+q_r)+\ell)$.
 \item If $c^{(i)}=0$, compute $w^{(i)}=\phi(g_2)^{b^{(i)}}$. 
 \item If $c^{(i)}=1$, compute $w^{(i)}=h\cdot \phi(g_2)^{b^{(i)}}$.
 \item Insert $\langle v^{(i)}, w^{(i)}, s^{(i)}, c^{(i)} \rangle$ in $\mathbb{T}$ and return $w^{(i)}$.
 \end{itemize}
 
 \item $\game_{3-4}$:We modify the signing algorithm $\Sign$ in Step 6 as follows:
 \begin{itemize}
 \item Set $v^{(0)} \leftarrow (\DID||\ord(\ADM))$, $v^{(j)} \leftarrow (\DID||m_j) \; (1\leq j \leq \#M)$.
 \item Query $v^{(j)}$ $(0 \leq j \leq \#M)$ to $\mathcal{O}^H$. We assume $\langle v^{(j)}, w^{(j)}, b^{(j)}, c^{(j)} \rangle$ to be the tuple in $\mathbb{T}$ for each $v^{(j)}$ $(1 \leq j \leq \#M)$.
 \item If $c^{(j)}=0$ $(0 \leq \forall j \leq \#M)$, go to Step \ref{signfix} of $\Sign$. Otherwise return $\perp$ and abort.
 \end{itemize}
 
  \item $\game_{3-5}$: We modify the signing algorithm $\Sign$ in Step \ref{signm}, \ref{signagg} as follows:
 \begin{itemize} 
  \item Compute $\sigma_{\ADM} \leftarrow \phi(u)^{b^{(0)}}\cdot \phi(g_2)^{r{b^{(0)}}}$.
  \item For all $m_j \in M$, compute $\sigma_{m_j} \leftarrow \phi(u)^{b^{(j)}}\cdot \phi(g_2)^{r{b^{(j)}}}$.
  \item Compute $\Sigma_{\agg} \leftarrow \sigma_{\ADM} \cdot \prod_{m_j \in M} \sigma_{m_j}$.
 \end{itemize}
 (By above modification, a signature $\Sigma_{\agg}$ can be generated without a knowledge of the $\skM$.)

 \item $\game_{3-6}$: We change a setting of $\mathcal{O}^{\Redact}$.
 \begin{itemize}
 \item Parse $\pk$ as $(\pkADM, \pkM, t, n)$ and $\sigma$ as $(\sigma_{\Fix}, \Sigma_{\agg})$.
 \item If $\DID \in \mathbb{L}^{u-1}$, then abort. 
 \item Set $\mathbb{L}^{u} \leftarrow \mathbb{L}^{u-1} \cup \{\DID\}$.
 \item If $\MOD \nsubseteq M \lor \MOD \cap \ADM \neq \emptyset$, then abort.
 \item Set $v^{(0)} \leftarrow (\DID||\ord(\ADM))$, $v^{(j)} \leftarrow (\DID||m_j) \; (1\leq j \leq \#M)$.
 \item Query $v^{(j)}$ $(0 \leq j \leq \#\MOD)$ to $\mathcal{O}^H$. 
 We assume $\langle v^{(j)}, w^{(j)}, b^{(j)}, c^{(j)} \rangle$ to be the tuple in $\mathbb{T}$ for each $v^{(j)}$ $(1 \leq j \leq \#\MOD)$.
 \item If $e(\sigma_{\Fix}, g_2) \neq e(w^{(0)}, \pkADM)$, then abort.
 \item If $e(\Sigma_{\agg}, g_2) \neq \prod_{0 \leq j \leq \#M}e(w^{(j)}, \pkM)$, then abort.
 \item If $c^{(j)}=0$ $(\forall m_j \in \MOD)$, go to next step. Otherwise return $\perp$ and abort.
 \item For all $m_j \in \MOD$, compute $\sigma_{m_j} \leftarrow \phi(u)^{b^{(j)}}\cdot \phi(g_2)^{r{b^{(j)}}}$.
 \item Compute $\sigma_{\MOD} \leftarrow \prod_{m_j \in \MOD} \sigma_{m_j}$, $\Sigma_{\agg}' \leftarrow \Sigma_{\agg} / \sigma_{\MOD}$.
 \item Set $M' \leftarrow M \backslash \MOD$, $\sigma' \leftarrow (\sigma_{\Fix}, \Sigma_{\agg}')$.
 \item Return $(M', \ADM, \DID, \sigma')$.
 \end{itemize}
 (Redactions can be done without the knowledge of the $\skM$.)

 \item $\game_{3-7}$: We receiving the output forgery $(M^*, \ADM^*, \DID^*, \sigma^*)$ from the adversary $\A_3$, 
 \begin{itemize}
 \item Set $v^{(0)} \leftarrow (\DID^*||\ord(\ADM^*))$, $v^{(j)} \leftarrow (\DID||m_j^*) \; (1\leq j \leq \#M^*)$.
 \item Query $v^{(j)}$ $(0 \leq j \leq \#M^*)$ to $\mathcal{O}^H$. We assume $\langle v^{(j)}, w^{(j)}, s^{(j)}, c^{(j)} \rangle$ to be the tuple in $\mathbb{T}$ for each $v^{(j)}$ $(0 \leq j \leq \#M^*)$.
 \item If $c^{(1)} = 1$ and $c^{(j)} = 0$ $(2 \leq j \leq \#M^*)$, then accept. Otherwise reject and abort. 
 \end{itemize}

\begin{lemma}\label{A3nnx}
If $H$ is the random oracle model, the following equation holds.
\begin{equation*}
\Adv_{\A_3}[\game_{3-3}] = \Adv_{\A_1}[\game_{3-2}]
\end{equation*}
\end{lemma}
Since the distribution of outputs of $\mathcal{O}^H$ in $\game_{3-3}$ and $\game_{3-2}$ is identical.

\begin{lemma} The following inequality holds.
\begin{equation*}
\Adv_{\A_3}[\game_{3-4}] \geq (1-1/((\ell+1)(q_s+q_r)+\ell))^{(\ell+1) q_s} \times \Adv_{\A_3}[\game_{3-3}].
\end{equation*}
\end{lemma}
Since the probability that each signing query does not abort at least \\$(1-1/((\ell +1 )(q_s+q_r)+(\ell+1)))^{(\ell+1)}$.

\begin{lemma} The following equation holds.
\begin{equation*} 
\Adv_{\A_3}[\game_{3-5}] = \Adv_{\A_3}[\game_{3-4}].
\end{equation*}
\end{lemma}
Since outputs of $\Sign$ in $\game_{3-5}$ and $\game_{3-4}$ are same.

\begin{lemma} The following inequality holds.
\begin{equation*}
\Adv_{\A_3}[\game_{3-6}] \geq (1-1/((\ell+1)(q_s+q_r)+\ell))^{(\ell+1) q_r} \times \Adv_{\A_3}[\game_{3-5}].
\end{equation*}
\end{lemma}
Since the probability that each redaction query does not abort at least \\$(1-1/((\ell +1 )(q_s+q_r)+\ell))^{(\ell+1)}$.

\begin{lemma}\label{A3gol} The following inequality holds.
\begin{equation*}
\begin{split}
\Adv_{\A_3}[\game_{3-7}] \geq& (1-1/((\ell+1)(q_s+q_r)+\ell))^{\ell-1}\\
&\;\;\;\times(1/((\ell+1)(q_s+q_r)+\ell)) \times \Adv_{\A_3}[\game_{3-6}].
\end{split}
\end{equation*}
\end{lemma}
Since an output $(M^*, \ADM^*, \DID^*, \sigma^*)$ satisfies $c^{(0)} = 0$.
The probability that $(M^*, \ADM^*, \DID^*, \sigma^*)$ satisfies $c^{(1)} = 1$ and $c^{(i)}=0$ $(2 \leq i \leq \#M^*)$ at least $(1-1/((\ell +1 )(q_s+q_r)+\ell))^{(\ell-1)} \times (1-1/((\ell +1 )(q_s+q_r)+\ell))$.

\end{itemize}
To summarize {\bf Lemma \ref{A2sta}}, {\bf Lemma \ref{A2nex}}, and {\bf Lemma \ref{A3nnx}} to {\bf \ref{A3gol}}, the following holds. (In the following equation, $e$ represents the Napier's constant.)
\begin{equation*}
\begin{split}
\Adv_{\A_3}[\game_{3-7}] &\geq  (1-1/((\ell+1)(q_s+q_r)+\ell))^{(\ell+1)(q_s+q_r)+\ell-1}\\
&\;\;\;\;\;\times (1/((\ell+1)(q_s+q_r)+\ell)) \times \Adv_{\A_3}[\game_{3-0}]\\
& \geq (1/e) \times (1/((\ell+1)(q_s+q_r)+\ell)) \times \Adv_{\A_3}[\game_{3-0}]
\end{split}
\end{equation*}

Now we construct the algorithm $\B_3$ which breaking the computational co-Diffie-Hellman assumption running the algorithm $\A_3$.
The operation of $\B_3$ for the input co-Diffie-Hellman problem instance $(g_2, g_2^{\alpha}, h^*)$ is changed to $h$ in $\game_{3-7}$ to $h^*$ and $u$ to $g_2^{\alpha}$.\\ 
Suppose $\B_3$ do not abort receiving a forgery $(M^*, \ADM^*, \DID^*, \sigma^*)$ from $\A_3$.
$\B_3$ parses $\sigma^*$ as $(\sigma_{\ADM^*}^*, \Sigma_{\agg}^*)$, sets $v^{(j)} \leftarrow ({\sf DID}^*||m_j^*) \; (1\leq j \leq \#M^*)$, and computes $w^{(1)} \leftarrow h\cdot \phi(u)^{b^{(1)}}\cdot \phi(g_2)^{r{b^{(1)}}}$,  $w^{(j)} \leftarrow \phi(u)^{b^{(j)}}\cdot \phi(g_2)^{r{b^{(j)}}} \;(2 \leq j \leq \#M^*)$.
Then $\B_3$ computes $\sigma^*_{m^*_1} \leftarrow \Sigma_{\agg}^*/ \prod^{\#M*}_{j=2} \sigma_{m_j}$.
Since $(M^*, \ADM^*, \DID^*, \sigma^*)$ is valid signature and $\pkM = g_2^{{\alpha}+r}$,
$e(\sigma^*_{m^*_1}, g_2) = e\left((w^{(1)})^{\alpha+r}, g_2\right)$ holds.
It implies that $\sigma^*_{m^*_1} = (w^{(1)})^{\alpha+r}=(h^* \cdot \phi(g_2)^{b^{(1)}})^{\alpha+r}$.
Threfore, $\B_3$ computes $(h^*)^{\alpha} = \sigma^*_{m^*_1} \cdot (\phi(u)^{b^{(1)}} \cdot (h^*)^r \cdot \phi(g_2)^{rb^{(1)}})^{-1}$ and outputs the solution $(h^*)^{\alpha}$ of the computational co-Diffie-Hellman problem instance $(g_2, g_2^{\alpha}, h^*)$.

Let $\epsilon_{{\sf co\mathchar`-cdh3}}$ is the probability that $\B_3$ breaks the computational co-Diffie-Hellman assumption.
We can bound the probability $\epsilon_{{\sf co\mathchar`-cdh3}} \geq \Adv_{\A_3}[\game_{3-7}]$ and $\epsilon_{{\sf co\mathchar`-cdh3}} \geq (1/e) \times (1/((\ell+1)(q_s+q_r)+ \ell)) \times \epsilon_{{\sf uf3}}$ holds. ($e$ represents the Napier's constant.)

Hence, if $\epsilon_{{\sf uf3}}$ is non-negligiable in $\lambda$, $\B_3$ breaks the computational co-Diffie-Hellman assumption with non-negligiable in $\epsilon_{{\sf co\mathchar`-cdh3}}$. \qed
\end{proof}

\section{Proof of Theorem \ref{ourtrans}} \label{ourtranspoof}

\begin{proof}
We proceed by a sequence of games.
Define $\Adv_{\A}[\game_{i}]$ as the advantage of an adversary $\mathcal{\A}$ in $\game_{i}$.

\begin{itemize}
 \item $\game_{0}$: Original transparency game in a redactable signature scheme. 
 \item $\game_{1}$: We change the redaction algorithm $\Redact$ in $\mathcal{O}^{\SR}$.  
 \begin{itemize}
 \item Skip the step 2 of $\RedInf$.
 \end{itemize}
\end{itemize}
Let $q_r$ be the total number of queries from an adversary $\A$ to $\mathcal{O}^{\Redact}$.
Then, $|\Adv_{\A}[\game_{1}] -  \allowbreak \Adv_{\A}[\game_{0}] | \leq q_{r} \times q_r/2^d$ holds.
We consider distribution of output $\mathcal{O}^{\SR}$ in case of $b=0$ and $b=1$ of $\game_{1}$.
Given an input $(M, \ADM, \MOD)$ to $\mathcal{O}^{\SR}$, $\mathcal{O}^{\SR}$ compute
\begin{itemize}
 \item $(M, \ADM, \DID_0, \sigma) \leftarrow \Sign(\sk, M, \ADM)$.
 \item $\RI_i \leftarrow \RedInf(\pk, \rk[i], M, \ADM, \DID_0, \sigma, \MOD) $ for $1 \leq i \leq n$.
 \item $(M', \ADM', \DID_0, \sigma_0) \leftarrow \ThrRed(\pk, M, \ADM, {\sf DID}_0, \sigma, \{\RI_i\}_{i=1}^n)$.
 \item $(M', \ADM', \DID_1, \sigma_1) \leftarrow \Sign(\sk, M',\ADM')$.
\end{itemize}
 
Distributions of ${\sf DID}_0$ and ${\sf DID}_1$ in $\game_{1}$ are identical and $\mathcal{O}^{\SR}$ skips the step 2 of $\RedInf$.  
Therefore, distributions of $\{(M', \ADM', \DID_0, \sigma_0)\}$ and $\{(M', \ADM', \DID_1, \sigma_1)\}$ outputted by $\mathcal{O}^{\SR}$ are identical. 
It means that $\Adv_{\A}[\game_{1}] = 1/2$.
Let $\epsilon_{\tran}$ is the advantage of an adversary $\mathcal{\A}$ in original $\game_{0}$.  
We can bound the probability $\epsilon_{\tran} \leq q_{r} \times q_r/2^d + 1/2$.
Therefore, our proposed $t$-out-of-$n$ redactable signature scheme in the one-time redaction model $\tnRSS$ $\Pi_1$ satisfies transparency. \qed
\end{proof}

\setcounter{tocdepth}{2}
\tableofcontents

\end{document}